\documentclass[]{aa}  

\usepackage{graphicx}
\usepackage{txfonts}
\usepackage{hyperref}
\usepackage{upgreek}
\usepackage{xcolor}
\usepackage{comment}
\usepackage{mathtools}
\usepackage{ulem}

\newcommand{\sectionname}{Sect.}
\newcommand{\dd}{\textrm{d}}

\newcommand{\deriv} [2] {\frac {\textrm{d} #1 } {\textrm{d} #2} }

\newcommand{\algn} [1] {
\begin{align} #1
\end{align}}

\let\originaleqref\eqref
\renewcommand{\eqref}{Eq.~\originaleqref}
\newcommand{\eq}[1] {Eq.\,(\ref{#1})}
\newcommand{\eqss}[2]{Eqs.~(\ref{#1})-(\ref{#2})}
\makeatletter
\newcommand{\eqs}[1]{%
    Eqs.~(\ref{#1})\checknextarg}
\newcommand{\checknextarg}{\@ifnextchar\bgroup{\gobblenextarg}{}}
\newcommand{\gobblenextarg}[1]{\@ifnextchar\bgroup{, (\ref{#1})\gobblenextarg}{ and (\ref{#1})}}
\makeatother

\begin{document} 


   \title{Evolution of the gravity offset of mixed modes in RGB stars}

  \titlerunning{Gravity offset of mixed modes in RGB stars} 
  
   \author{C. Pin\c con\inst{1,2,3}, M. Takata\inst{3,4} and B. Mosser\inst{3}
          }
          
  \authorrunning{C. Pin\c con et al.}
   \institute{Institut d’Astrophysique Spatiale, Univ. Paris-Sud, CNRS, Universit\'e Paris-Saclay,
B\^atiment 121, 91405 Orsay CEDEX, France
         \and
         STAR Institute, Université de Liège, 19C Allée du 6 Août, B-4000 Liège, Belgium\\
         \email{charly.pincon@uliege.be}
         \and
             LESIA, Observatoire de Paris, PSL Research University, CNRS, Sorbonne Universit\'es, Univ. Paris Diderot, 5 place Jules Janssen, 92195 Meudon, France
           \and
                Department of Astronomy, School of Science, The University of Tokyo, 7-3-1 Hongo, Bunkyo-ku, 113-0033 Tokyo, Japan
             }

   \date{\today}

  \abstract
   {Observations of mixed modes in evolved low-mass stars enable us to probe the properties of not only the outer envelope of these stars, but also their deep layers. Among the seismic parameters associated with mixed modes, the gravity offset, denoted with $\varepsilon_{\rm g}$, is expected to reveal information on the boundaries of the inner buoyancy resonant cavity. This parameter was recently measured for hundreds of stars observed by the {\it Kepler} satellite and its value was shown to change during evolution.}
   {In this article, we theoretically investigate the reasons for such a variation in terms of structure properties, focusing only on the red giant branch.}
   {Using available asymptotic analyses and a simple model of the Brunt-Väisälä and Lamb frequencies in the upper part of the radiative zone, we derived an analytical expression of $\varepsilon_{\rm g}$ for dipolar modes and compared its predictions to observations.}
   { First, we show that the asymptotic value of $\varepsilon_{\rm g}$ well agrees with the mean value observed at the beginning of the ascent of the red giant branch, which results from the high density contrast between the helium core and the base of the convective envelope. Second, we demonstrate that the predicted value of the gravity offset also explains the sharp decrease in $\varepsilon_{\rm g}$ observed for the more luminous red giant stars of the sample. This rapid drop turns out to occur just before the luminosity bump and result from the kink of the Brunt-Väisälä frequency near the upper turning point associated with the buoyancy cavity as stars evolve and this latter becomes close to the base of the convective envelope. The potential of $\varepsilon_{\rm g}$ to probe the value and slope of the Brunt-Väisälä frequency below the base of the convective region is clearly highlighted.}
   {The observed variation in $\varepsilon_{\rm g}$ and its link with the internal properties on the red giant branch are now globally understood. This work motivates further analyses of the potential of this parameter as a seismic diagnosis of the region located between the hydrogen-burning shell and the base of the convective envelope, and of the local dynamical processes associated for instance with core contraction, the migration of the convective boundary, or overshooting.}


   \maketitle


%
\section{Introduction}

The space-borne missions CoRoT \citep[e.g.,][]{Baglin2006a,Baglin2006b} and {\it Kepler} \citep[e.g.,][]{Borucki2010} now allow access to seismic data for thousands of stars, from the main sequence to the core helium burning stages, and even to white dwarfs. These additional constraints have led to a revolution in stellar physics and continue to reveal the dynamics at work in the inner layers of these stars \citep[e.g.,][]{Chaplin2013,DiMauro2016}.

One of the important advances in the field brought by these data results from the detection and study of mixed modes in evolved stars \citep[e.g.,][]{Mosser2016,Hekker2017}. Mixed modes constitute a peculiar type of eigenmodes that oscillate in a central cavity, or buoyancy cavity, where they behave as gravity modes, and an external cavity, or acoustic cavity, where they behave as pressure modes; both of these cavities are coupled by an evanescent region located between the profiles of the Brunt-Väisälä and Lamb frequencies (see \figurename{}~\ref{prop_diag}). Assuming that the short-wavelength WKB approximation is met in each cavity (i.e., in the asymptotic limit), their oscillation frequencies satisfy the general quantization relation \citep[][]{Shibahashi1979,Tassoul1980,Takata2016a,Takata2016b}
\algn{
\cot \Theta_{\rm g} \tan \Theta_{\rm p} = q \; ,
\label{asymptotic relation}
}
where $\Theta_{\rm g}$ and $\Theta_{\rm p}$ are frequency dependent phase terms associated with the propagation of waves in the buoyancy and the acoustic cavities, respectively, and $q$ is the coupling factor that quantifies the level of interaction between both cavities through the intermediate evanescent region.

As shown by \eq{asymptotic relation}, the mixed mode frequency pattern contains information on various regions of evolved stars and in particular on their deepest layers. First, observations of mixed modes have led to the measurement, for thousands of stars, of the dipolar period spacing \citep[e.g.,][]{Bedding2011,Mosser2012b,Vrard2016}, the value of which is sensitive to both the properties of the helium core and the evolutionary stage \citep[e.g.,][]{Montalban2013,Mosser2014,Lagarde2016}. The coupling factor of dipolar modes has also been measured for such a large sample of stars and has been shown to vary during evolution because of the structural changes of the region located between the hydrogen-burning shell and the base of the convective envelope \citep[e.g.,][]{Mosser2017b,Hekker2018}. Additionally, the rotational splittings of mixed modes have provided important constraints on the redistribution of internal angular momentum during the post-main sequence stage \citep[e.g.,][]{Deheuvels2014,Gehan2018}.

More recently, several works have focused on a new seismic parameter associated with mixed modes, i.e., the gravity offset \smash{denoted with $\varepsilon_{\rm g}$} \citep[e.g.,][]{Buysschaert2016,Hekker2018,Mosser2018}. This parameter on the order of unity, which appears inside $\Theta_{\rm g}$ in \eq{asymptotic relation}, mainly depends on the properties of the regions close to boundaries of the buoyancy cavity. \cite{Mosser2018} were able to measure the gravity offset with precision for hundreds of stars on the red giant branch (hereafter, RGB) and on the red clump. The authors showed that the observed value remains about constant at the beginning of the RGB and then rapidly drops before the stars reach the luminosity bump. For more evolved red clump stars, the spread in the measurements turns out to be much larger than for RGB stars.

In this work,
we aim at interpreting the evolution of the gravity offset on the RGB from a theoretical point of view in terms of internal structure. For this purpose, the available asymptotic analyses of dipolar mixed modes are considered. The subsequent statements thus assume that the oscillation wavelength is much smaller than the structure scale height in both resonant cavities (i.e., in the WKB approximation). The paper is organized as follows. In \sectionname{}~\ref{setting}, we properly introduce the notion of gravity offset and clearly pave the way for the present investigation. A general physical interpretation of this parameter within the framework of the asymptotic limit is then given in \sectionname{}~\ref{asymptotic epsilon_g}. Using this theoretical background and simple models of the Brunt-Väisälä and Lamb frequencies, we derive in \sectionname{}~\ref{evolution on RGB} an analytical expression for $\varepsilon_{\rm g}$ that explicitly accounts for the structural changes occurring during evolution. This model is subsequently used to explain and interpret the observations by \cite{Mosser2018}.
We discuss the results in \sectionname{}~\ref{discussion} and \sectionname{}~\ref{conclusions} is devoted to conclusions.

%
\section{Setting the stage}
\label{setting}

In this introductory section, we expose the problem and clarify the underlying goals of the present investigation. In all the following, we implicitly focus on dipolar mixed modes. Indeed, among mixed modes that penetrate into the stellar core, only dipolar modes can currently bring us information on the innermost properties of evolved stars since they have the highest amplitude at the surface \citep[e.g.,][]{Dupret2009,Grosjean2014}.

\subsection{Empirical introduction of the gravity offset}
\label{observational definition}

Before any theoretical study on the subject, the notion of gravity offset was first introduced in an observational context by \citet{Mosser2012a}. To fit the asymptotic expression of mixed modes in \eq{asymptotic relation} to real frequency spectra, \cite{Mosser2012a} parametrized the phase $\Theta_{\rm g}$ associated with the buoyancy cavity as
\algn{
\Theta_{\rm g}= \pi \left( \frac{1}{\nu \Delta \Pi_1^{\rm obs}}-\varepsilon_{\rm g}^{\rm obs}\right) \; ,
\label{Theta_g obs}
}
where $\nu$ is the oscillation cyclic frequency (with $\sigma = 2\pi \nu$ the angular frequency), \smash{$\Delta \Pi_1^{\rm obs}$} is the observed dipolar period spacing, and \smash{$\varepsilon_{\rm g}^{\rm obs}$} is the so-called observed gravity offset. Within the framework of the fitting procedure, \smash{$\Delta \Pi_1^{\rm obs}$} and \smash{$\varepsilon_{\rm g}^{\rm obs}$} were supposed to be frequency independent in the observed narrow frequency range around $\nu_{\rm max}$, representing the frequency at maximum oscillation power (see \eq{nu_max} for an expression). 

The presence of the dominant first term in the brackets of the right-hand side of \eq{Theta_g obs} as well as its frequency dependence on $1/\nu$ were motivated by the results of the asymptotic analyses of mixed modes by \cite{Shibahashi1979} and \cite{Tassoul1980}. This term is close to the number of radial nodes of a mixed mode in the buoyancy cavity, denoted with \smash{$n_{\rm g}$}, which is very large in the asymptotic limit. In contrast, the small term on the
order of unity, $\varepsilon_{\rm g}^{\rm obs}$, was introduced for empirical reasons. Indeed, \cite{Mosser2012a} found that the fits of the asymptotic relation to real data were better with non-null values of \smash{$\varepsilon_{\rm g}^{\rm obs}$}, in particular in subgiant and young red giant stars. Given its role in the asymptotic relation and the $\pi$-periodic behavior of the cotangent function in \eq{asymptotic relation}, the value of \smash{$\varepsilon_{\rm g}^{\rm obs}$} can be measured only modulo unity. A theoretical justification of the empirical expression of $\Theta_{\rm g}$ in \eq{Theta_g obs} is therefore needed to give a clear physical
meaning to the observed values of the gravity offset.

\subsection{Unclear previous interpretations}
\label{unclear}

The gravity offset of mixed modes has been interpreted through an analogy with the asymptotic analyses of pure gravity modes provided by \cite{Provost1986} in the case of the Sun, or by \cite{Brassard1992} in the case of ZZ Ceti stars. Nevertheless, the link between the results of these works and the case of mixed modes in evolved stars deserves to be clarified because of different evolutionary stages and structural properties as well as types of oscillation modes.

\begin{figure}
\centering
\includegraphics[scale=0.35,trim= 0.5cm 0cm 2cm 1cm, clip]{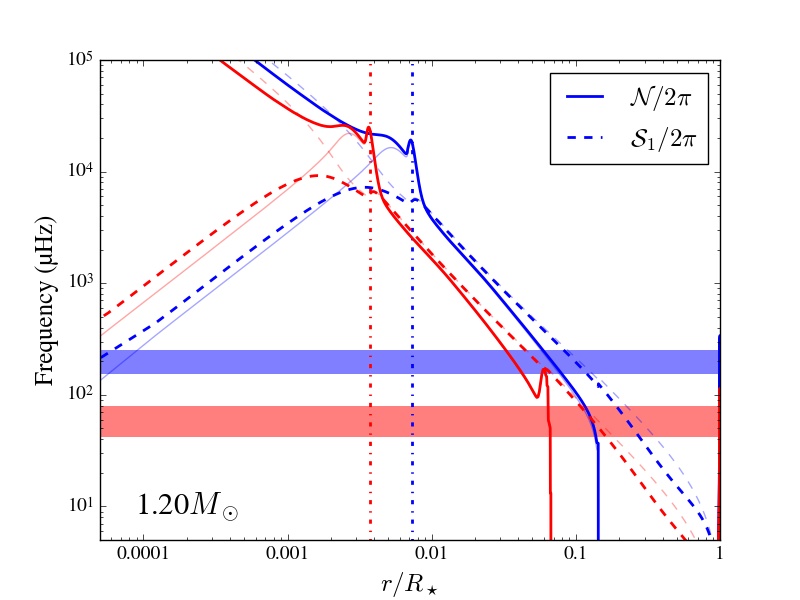} 
\caption{Profile of the modified Brunt-Väisälä and $\ell=1$ Lamb frequencies (thick solid and dashed lines, respectively) for two 1.2$M_\odot$ red giant models with $\nu_{\rm max} \approx 200 ~\mu$Hz (blue) and $ \nu_{\rm max} \approx 60 ~\mu $Hz (red). These both models correspond to stars at the beginning of the ascent of the RGB and at a point just before the luminosity bump, respectively. The usual Brunt-Väisälä and $\ell=1$ Lamb frequencies, $N$ and $S_1$, are represented by the light thin solid and dashed lines, respectively. The vertical dash-dotted lines indicate the hydrogen-burning shell and the horizontal strip symbolizes a typical $6\Delta \nu$-wide observed frequency range around $\nu_{\rm max}$, for each model with the corresponding color. The buoyancy and acoustic cavities for a mode of frequency $\nu$ are located where $2 \pi \nu \le (\mathcal{N}~{\rm and}~ \mathcal{S}_1$) and $2 \pi \nu \ge (\mathcal{N}~{\rm and}~\mathcal{S}_1)$, respectively.} 
\label{prop_diag}
\end{figure}

Actually, the ambiguity also extends to the physical interpretation of the observed values of the period spacing, as already discussed in \citet[][cf. Sect.~4.4 of this paper]{Takata2016a}. Indeed, previous works often considered \smash{$\Delta \Pi_1^{\rm obs}$} as equal to the asymptotic quantity \smash{$\widetilde{\Delta \Pi_1}$}, which appears in the expression of $\Theta_{\rm g}$ derived by \cite{Shibahashi1979} and \cite{Tassoul1980}, and is defined as
\begin{align}
\widetilde{\Delta \Pi_1} \equiv \frac{2\pi^2}{\sqrt{2}} \left( \int_{r_1(\sigma)}^{r_2(\sigma)} \frac{N}{r}\dd r\right)^{-1} \; ,
\label{Delta Pi tilde}
\end{align}
where $r$ is the radius in the star and $N$ is the usual Brunt-Väisälä frequency \citep[e.g.,][for a definition]{Unno1989}. The bounds of the integral, $r_1$ and $r_2$, correspond to the inner and the upper turning points associated with the buoyancy cavity. In the formulations of \cite{Shibahashi1979} and \cite{Tassoul1980}, both represent the physical boundaries of the cavity and are equal to the radii at which $\sigma = N(r)$. As an illustration, the radial profile of $N$ is shown in \figurename{}~\ref{prop_diag} for two 1.2$M_\odot$ red giant models computed with the stellar evolution code CESTAM \citep[][]{Marques2013}. 
As input physics, the chemical composition corresponds to a solar mixture, as provided by \cite{AGS2009}, which has solar-calibrated initial helium and metal abundances $Y_0=0.25$ and $Z_0=0.013$. The OPAL 2005 equation of state and the NACRE nuclear reaction rates were considered. Microscopic diffusion, overshooting, and rotation were neglected and the stellar atmosphere was constructed following the Eddington gray approximation. The mixing-length theory of convection was used with a solar-calibrated parameter $\alpha_{MLT}=1.65$. Both models chosen as examples correspond to stars at the beginning of the ascent of the RGB with $\nu_{\rm max} \approx 200 $ and just before the luminosity bump with $\nu_{\rm max} \approx 60 ~\mu $Hz, respectively. We note that the radial profiles of the critical frequencies as well as their evolution with time in stars with masses between $1M_\odot$ and $2M_\odot$ follow the same trend between the beginning of the RGB and the luminosity bump as that depicted in \figurename{}~\ref{prop_diag}. Indeed, whatever their mass in this range, red giant stars evolve along the Hayashi line in the Herztsprung-Russell diagram and thus share similar properties \citep[e.g.,][]{Kippenhahn2012}.
In \figurename{}~\ref{prop_diag}, a typical $6\Delta \nu$-wide observed frequency range around $\nu_{\rm max}$, as expected from theoretical estimates and observations in RGB stars \citep[e.g.,][]{Grosjean2014,Mosser2018}, is indicated for each model, where $\Delta \nu$ is the large frequency separation between two consecutive acoustic radial modes. Since both turning points in the buoyancy cavity are located where $\sigma=N$ in the asymptotic formulation of \cite{Shibahashi1979} and \cite{Tassoul1980}, the variation of a given turning point $r_i$ with the oscillation frequency is measured by
\algn{
\deriv{\ln r_i}{\ln \sigma} = \left(\deriv{\ln N}{\ln r} \right)^{-1}_{r_i} \; .
\label{drds}
}
As $(\dd \ln N/\dd \ln r)$ is finite in the central regions of each model considered in \figurename{}~\ref{prop_diag}, the inner turning point $r_1$ depends on $\sigma$ according to \eq{drds}. Similarly, the outer turning point $r_2$ is also frequency dependent for the young red giant model with $ \nu_{\rm max} \approx 200 ~\mu $Hz. As a consequence, \smash{$\widetilde{\Delta \Pi_1}$} in \eq{Delta Pi tilde} is frequency dependent as well. It is therefore clear that this quantity is not representative of the observed period-spacing \smash{$\Delta \Pi_1^{\rm obs}$} as this latter is supposed to be frequency independent. We also anticipate that the outer turning point $r_2$ is, in contrast, independent of frequency in the more evolved model with $ \nu_{\rm max} \approx 60 ~\mu $Hz considered in \figurename{}~\ref{prop_diag}.

Since the observed gravity offset and observed period spacing of mixed modes are related to each other by \eq{Theta_g obs}, we need to understand simultaneously the link of both observed parameters with the asymptotic expressions to achieve our purpose. To do so, we have to consider a more detailed examination of the available asymptotic analyses of mixed modes.
A similar study was performed by \cite{Mosser2013b} in the case of the radial acoustic modes observed in red giant stars. Indeed, for these modes, the authors showed that the observed value of the frequency large separation and the asymptotic value are not equivalent. They then demonstrated the need to consider a second-order expansion around $\nu_{\rm max}$ of the asymptotic expression of radial modes derived by \cite{Tassoul1980} to make the link between both
and theoretically interpret the observed values. A similar approach is used in this work in the case of mixed modes, as presented in the next section.

\subsection{Local asymptotic definitions around $\nu_{\rm max}$}
\label{asymptotic definition}

The comparison of any asymptotic expressions of the gravity offset and of the period spacing with the observed values are only relevant if they have equivalent definitions. This can be met if the analytical expression of $\Theta_{\rm g}$ is proved to have a similar form to \eq{Theta_g obs} in terms of frequency dependence over the observed frequency range, at least with a level of error lower than the uncertainties on the measurements of the gravity offset denoted with $\sigma_{\varepsilon}$. Nevertheless, such an identification is not straightforward since we have seen that $\Theta_{\rm g}$ can depend on the frequency in a more complicated way than \eq{Theta_g obs}; see the frequency dependence of the bounds of the integral in \eq{Delta Pi tilde} for instance.

Actually, whatever the expression of $\Theta_{\rm g}$, a theoretical justification for the form given in \eq{Theta_g obs} can be obtained through a local analysis over a narrow frequency range around $\nu_{\rm max}$, in an analogous way to the case of the radial acoustic modes in red giant stars \citep{Mosser2013b}. Indeed, to demonstrate this, we formally rewrite $\Theta_{\rm g}$ as
\algn{
\Theta_{\rm g} \equiv \frac{F(\nu)}{\nu} \; ,
\label{Theta_g formal}
}
where $F$ is a differentiable function of frequency. Using a first-order Taylor expansion of $F$ around $\nu=\nu_{\rm max}$, we obtain
\algn{
F(\nu)=F(\nu_{\rm max})+F^\prime(\nu_{\rm max}) (\nu -\nu_{\rm max}) + G(\nu)(\nu -\nu_{\rm max})\; ,
\label{Taylor F}
}
where $(^\prime)$ denotes the derivative with respect to frequency. In the last equation, $G(\nu)$ is a real function (not necessarily differentiable) such as $G(\nu) \rightarrow 0$ as $\nu \rightarrow \nu_{\rm max}$ and that encapsulates all the higher order terms. In a small enough frequency range around $\nu_{\rm max}$, it is always possible to assume that the last term in the right-hand side of \eq{Taylor F} is negligible compared to the preceding ones. Substituting the truncated first-order expansion of \eq{Taylor F} into \eq{Theta_g formal}, $\Theta_{\rm g}$ is thus about equal around $\nu_{\rm max}$ to
\algn{
\Theta_{\rm g} \approx \frac{F(\nu_{\rm max})-\nu_{\rm max} F^\prime (\nu_{\rm max})}{\nu} + F^\prime(\nu_{\rm max})  \; ,
\label{theta_g taylor}
}
which has the same form as \eq{Theta_g obs} with respect to the frequency dependence. According to \eq{theta_g taylor}, the local asymptotic dipolar period spacing and gravity offset around $\nu_{\rm max}$, denoted with $\Delta \Pi_1$ and $\varepsilon_{\rm g}$ in the following, can be identified, respectively,
 as
\algn{
\Delta \Pi_1 =  \frac{\pi}{F(\nu_{\rm max})-\nu_{\rm max} F^\prime (\nu_{\rm max})}
\label{local Delta Pi}
}
and
\algn{
\varepsilon_{\rm g}= -\frac{F^\prime(\nu_{\rm max})}{\pi} \; ,
\label{local epsilon_g}
}
so that $\Theta_{\rm g}$ can now be approximated around $\nu_{\rm max}$ as
\algn{
\Theta_{\rm g} \approx \pi \left( \frac{1}{\nu \Delta \Pi_1}-\varepsilon_{\rm g} \right) \; ,
\label{theta_g theor}
}
as proposed before through empirical considerations. We see a posteriori that a first-order Taylor expansion of the $F$ function is sufficient in \eq{Taylor F} to reproduce the empirical form in \eq{Theta_g obs}. Nevertheless, we emphasize that the identification in \eqs{local Delta Pi}{local epsilon_g}~remains accurate if and only if the considered frequency range around $\nu_{\rm max}$ is small enough for the higher order terms in \eq{Taylor F} to be negligible. Otherwise, the asymptotic gravity offset would be frequency dependent and could not be representative of the fitted values in the considered frequency range.

The question that naturally follows in practice is thus to know whether this relation is valid over all the observed frequency range around $\nu_{\rm max}$ or not with a level of error lower than $\sigma_{\varepsilon}$. According to \eqs{Theta_g formal}{Taylor F}, this is met if the remainder around $\nu_{\rm max}$,
\algn{
\gamma (\nu;\nu_{\rm max}) \equiv \frac{G(\nu)(\nu-\nu_{\rm max})}{\pi \nu }\; ,
\label{gamma}
}
is much smaller than $\sigma_{\varepsilon}$ over the observed frequency range.
Therefore, the comparison between the local asymptotic values and the observed values of both parameters are relevant if this latter condition is satisfied. Otherwise, we must conclude that the physical modeling and hypotheses on which the proposed asymptotic expression of $\Theta_{\rm g}$ is based are insufficient to interpret the observations. This condition is checked before any comparison in the following (see Sects.~\ref{as eps}~and~\ref{validity local}).

%
\section{Asymptotic interpretation of the gravity offset}
\label{asymptotic epsilon_g}

After having set the general context and goals of the present study, we now discuss in more detail the physical meaning of the gravity offset within the framework of the asymptotic limit. All the theoretical background useful for the present investigation is introduced in this section.

\subsection{General physical origin}

\label{general origin}

In a first step, it is instructive to discuss the physical origin of the gravity offset from a general point of view using simple considerations. To do so, considering the formulation of mixed modes proposed by \cite{Takata2016b} appears judicious. Using a progressive-wave picture of the oscillations in both cavities and basic wave principles, \cite{Takata2016b} retrieved the asymptotic quantization condition in \eq{asymptotic relation} and expressed $\Theta_{\rm g}$ in the very general form\footnote{The quantization condition of mixed modes is provided by \smash{Eqs.~(15)-(19)} of \cite{Takata2016a}, where the variable $X_{\rm G}$ is equal to the wavenumber integral over the buoyancy cavity.}
\algn{
\Theta_{\rm g}=\int_{r_1}^{r_2} \mathcal{K}_r \dd r~-\frac{\theta_{\rm G}}{2} -\frac{\delta}{2}+\frac{\pi}{2} \; ,
\label{theta_g_2}
}
where $\mathcal{K}_r$ is the radial wavenumber, which takes a null value at $r_1$ an $r_2$, and where $\theta_{\rm G}$ and $\delta$ are the phase lags\footnote{The phase lags are defined as the argument of the amplitude ratio of the reflected plane wave to the incident plane wave, which are associated with one given wave dependent variable. In this formulation, the direction of the propagation is determined by that of the propagation of the wave energy, that is of the group velocity.} introduced after the reflection of an incident wave propagating in the buoyancy cavity toward $r_1$ and $r_2$, respectively, with values in $[0,2\pi]$.

On the one hand, the comparison between \eqs{theta_g theor}{theta_g_2}~shows that a first contribution to the gravity offset may result from $\theta_{\rm G}$ and $\delta$, the phase lags introduced at the reflection near the center and the intermediate evanescent region, respectively. These terms, on the order of unity, seem missing at first sight in the asympotic expression of mixed modes derived by \cite{Shibahashi1979} under the Cowling approximation \citep[e.g.,][]{Cowling1941}. Actually, this results from the fact that $\delta + \theta_{\rm G} \approx \pi$ in this formulation. Indeed, \cite{Shibahashi1979} first demonstrated that the wave function (i.e., the dependent variable $\varv$ of his paper) near the innermost (single) turning point $r_1$ takes the form of an Airy function of the first kind, so that the phase lag introduced at the reflection in $r_1$ is equal to $\theta_{\rm G}=+\pi/2$ \citep[e.g., see also][]{Brekhovskikh1980}\footnote{Unlike a phase loss of $-\pi/2$ found by \cite{Brekhovskikh1980}, we obtain a phase gain of $+\pi / 2$ since, in the case of gravity waves, the direction of the propagation of the wave energy is in the opposite direction of the phase velocity \citep[e.g.,][]{Unno1989}.}. Secondly, the analysis of \cite{Shibahashi1979} assumed a very thick evanescent region, and therefore the same wave function also takes the form of an Airy function of the first kind close to the outer turning point $r_2$ and $\delta \approx + \pi /2$ for the same reasons as mentioned above. As a result, the sum of both phase lags is equal to $\pi$
in the hypothesis of a thick evanescent region and the last three terms in \eq{theta_g_2} cancel out. \cite{Takata2016a} showed later that this does not hold true when the evanescent region is very thin in red giant stars or when the influence of the perturbation of the gravitional potential is taken into account (i.e., in the non-Cowling approximation), as we see in the subsequent sections.

On the other hand, \eq{theta_g_2} also shows that another contribution to $\varepsilon_{\rm g}$ may come from the wavenumber integral, and in particular, from the integral over the region in the vicinity of the turning points. Indeed, far enough away from the turning points, that is in the middle part of the buoyancy cavity, $\mathcal{K}_r$ is provided in a good approximation by \citep[e.g.,][]{Unno1989}
\algn{
\mathcal{K}_r \approx \frac{\sqrt{2}}{r} \frac{N(r)}{2\pi \nu} \; .
\label{K_r approx}
}
This approximation is in fact valid when $2\pi \nu \ll N$, which is met in the middle of the buoyancy cavity for the frequency range observed in typical red giant stars (e.g., see \figurename{}~\ref{prop_diag}). Consequently, the wavenumber integral over the middle part of the buoyancy cavity mostly contributes to the leading-order term in $\Theta_{\rm g}$ that is proportional to $1/\nu$. In contrast, near the turning points, the approximation in \eq{K_r approx} is not valid since $\mathcal{K}_r$ tends to zero. There, say close to a turning point $r_i$, it is preferable to assume that $\mathcal{K}_r$ approximately takes the simple form
\algn{
\mathcal{K}_r \approx K(r_i) \left |\frac{r-r_i}{\Delta r (r_i)}\right|^{\alpha(r_i)}~~~~\mbox{for}~~~~\left|r -r_i  \right| \lesssim \Delta r (r_i)\; ,
}
where $K(r_i)$, $\alpha(r_i)$ and $\Delta r(r_i)$ are positive real quantities depending on the properties of the medium near $r_i$, with $\Delta r(r_i)$ chosen such as $\mathcal{K}_r (r) \Delta r (r_i) \lesssim 1$ in the considered region. In these considerations, the wavenumber integral over the region where $|r -r_i | \lesssim \Delta r (r_i)$ is thus provided in a first approximation by
\algn{
\pm \int_{r_i}^{r_i\pm \Delta r} \mathcal{K}_r \dd r\approx  \frac{\mathcal{K}_r(r_i+\Delta r) \Delta r (r_i)}{\alpha(r_i)} \; .
}
This simple estimate demonstrates, first, that the contribution of the wavenumber integral in the vicinity of the turning points to the total integral depends on the properties of the region close to $r_i$. Second, this estimate also shows that it is on the
order of unity at most since in general $\alpha(r_i)\gtrsim 1$ and $\mathcal{K}_r(r_i+\Delta r) \Delta r (r_i) \lesssim 1$ by definition.

The value of the gravity offset is therefore related to the properties of the boundaries of the buoyancy cavity (i.e., close to $r_1$ and $r_2$), through the contributions of the phase lags introduced at reflection and of the wavenumber integral in the vicinity of the turning points. As expected from the observational definition, both contributions are on the
order of unity at most. In contrast, they turn out to depend on the oscillation frequency since the turning point is in general frequency dependent. This emphasizes the need for a local asymptotic analysis around $\nu_{\rm max}$, that is such as proposed in \sectionname{}~\ref{asymptotic definition} to disentangle the contributions to each term in the first-order expansion of $\Theta_{\rm g}$ as a function of frequency that is given in \eq{theta_g taylor}, and thus interpret the observed values. At this stage, the computation of $\Delta \Pi_1$ and $\varepsilon_{\rm g}$ still needs the knowledge of the expression of $\Theta_{\rm g}$ in \eq{theta_g_2} for typical red giant stars.

\subsection{Link with the structure properties in red giant stars}
\label{link RGB}

All the physical information in \eq{theta_g_2} is encapsulated in the quantities $\theta_{\rm G}$, $\delta$, and $\mathcal{K}_r$. To go further, the link between these quantities and the internal properties is commonly made through more detailed analyses of the stellar oscillation equations considering realistic stellar structure.
We discuss and quantify these terms in the light of the asymptotic analysis by \citet[][hereafter, T16]{Takata2016a} for dipolar mixed modes in red giant stars. Regarding $\varepsilon_{\rm g}$, this latter work brought two main important theoretical improvements compared to the analyses of \cite{Shibahashi1979} and \cite{Tassoul1980}. First, it fully accounts for the effect of the perturbation of the gravitational potential on oscillations \citep[][]{Takata2006a}. Second, it considers the limiting case of a very thin evanescent zone. Nevertheless, the hypothesis of a very thick evanescent zone, as usually assumed by previous works \citep[][]{Shibahashi1979,Tassoul1980}, can be easily retrieved from the result of T16, and thus using this formulation enables us to discuss both cases in a convenient way. In the subsequent paragraphs, we note the main results of T16 concerning the gravity offset but present these in a different way to explicitly make the link with $\theta_{\rm G}$, $\delta,$ and $\mathcal{K}_r$.

Near the center of the star, the analysis is made very general because of the well-known behaviors of the internal structure and wave-displacement vector that are imposed by the central boundary conditions. First, the treatment of T16 in this region (cf. \sectionname{}~3.3 of T16) is similar to that of \cite{Shibahashi1979} except for an essential difference. Indeed, owing to the influence of the perturbation of the gravitional potential, the behavior of the dependent variable $Y_1$ in T16, which is equivalent to the dependent variable $\varv$ of \cite{Shibahashi1979} in the limit of the Cowling approximation, differs from an Airy function of the first kind near the innermost turning point $r_1$ (see \sectionname{}~\ref{general origin}). As a consequence, the phase lag introduced at the reflection in $r_1$ for the variable $Y_1$ turns out to be decreased by $\pi$ compared to its value in the Cowling approximation; thus it is equal in this case to (modulo $2\pi$)
\algn{
\theta_{\rm G}=-\pi/2 \; .
\label{theta_G}
}
For sake of simplicity, the basic derivation of $\theta_{\rm G}$ is detailed in \appendixname{}~\ref{phase lag}.
Second, the analysis of T16 (cf. \sectionname{}~3.3 and \appendixname{}~4 of T16) also accounted for the singularity at the center, which can impact both the location of the innermost turning point $r_1$ and the wavenumber integral over this region and therefore the value of $\varepsilon_{\rm g}$. The T16 author finally found 
\algn{
\int_{r_1}^{\tilde{r}} \mathcal{K}_r \dd r = \frac{\sqrt{2}}{\sigma} \int_0^{\tilde{r}} \frac{N}{r} \dd r -\frac{5\pi}{4} + O\left(\frac{\sigma}{N(\tilde{r})}\right)\; ,
\label{int centre}
}
where $N$ is the usual Brunt-Väisälä frequency and the big $O$ Bachmann-Landau notation has been used. In \eq{int centre}, $\tilde{r}$ is an arbitrary radius inside the buoyancy cavity that is far enough away from the turning points for the asymptotic limit to be valid and the residual to be negligible (i.e., $\sigma/N(\tilde{r}) \ll 1$).

In the upper part of the buoyancy cavity (i.e., for \smash{$\tilde{r} \le r \le r_2$)}, T16 (cf. \sectionname{}~3.2 of T16) simultaneously treated both closely-related turning points associated with the evanescent region. As a result, the same wave function (i.e., $Y_1$) takes this time the form of a linear combination of the Weber functions and the phase lag introduced at $r_2$ is given from the argument of Eq.~(99) of T16\footnote{The argument of Eq.~(99) of T16 is equal to the phase lag between the dependent variables $W_1$ and $W_2$ of T16, which is directly related to the phase lag between the incident and reflected components of $Y_1$, as shown by Eq.~(32) of T16.}, that is equal to
\algn{
\delta = 2\Psi+\pi\; ,
\label{delta}
}
where the function $\Psi$ is given in Eqs.~(85)-(86) of T16 and depends on the properties of the evanescent region. In the case of a very thick evanescent region, we have $\Psi \rightarrow - \pi/4$, which implies $\delta\rightarrow +\pi/2$, in agreement with the discussion in \sectionname{}~\ref{general origin}. 

Finally, to compute the expression of $\Theta_{\rm g}$, we also have to provide the expression of the radial wavenumber in the upper part of the buoyancy cavity. In the WKB approximation, it is equal to (T16)
\algn{
\mathcal{K}_r=\frac{\sqrt{2} J}{r} \frac{\sigma}{\mathcal{S}_1} \left( \frac{\mathcal{N}^2}{\sigma^2}-1\right)^{1/2} \left(\frac{\mathcal{S}_1^2}{\sigma^2}-1 \right)^{1/2} \; ,
\label{K_r}
}
where $\mathcal{N}$ and $\mathcal{S}_1$ are the modified Brunt-Väisälä and $\ell=1$ Lamb frequencies, respectively, which account for the perturbation of the gravitational potential. Both are related to the usual Brunt-Väisälä and Lamb frequencies, $N$ and $S_1$ \citep[see, e.g.,][for definitions]{Unno1989}, through the equalities $\mathcal{N}=N/J$ and $\mathcal{S}_1=S_1 J$, where $J=1-\rho(r)/\rho_{\rm av}(r)$, and $\rho$ is the local density and $\rho_{\rm av}$ is the mean density in the sphere of radius $r$. As an illustration, their profiles are shown in \figurename{}~\ref{prop_diag}. This figure shows that the difference between the modified and usual critical frequencies is large only below the hydrogen-burning shell and is negligible in the outer layers. Moreover, we can see that the turning points associated with the buoyancy cavity turn out to satisfy $\sigma = \mathcal{S}_1(r_1)$ and $\sigma = \mathcal{N}(r_2)$ in the non-Cowling case. The frequency dependence of the turning points in this case is thus also measured by \eq{drds}, but by replacing $N$ in this latter equation with $\mathcal{S}_1$ and with $\mathcal{N}$ for $r_1$ and $r_2$, respectively. Nevertheless, given the similar behaviors of $N$ and $\mathcal{S}_1$ close to $r_1$ and of $N$ and $\mathcal{N}$ close to $r_2$, the same conclusions as in the Cowling approximation hold valid about the frequency dependence of $r_1$ and $r_2$. The ambiguity on the physical interpretation of the period spacing and of the gravity offset that was highlighted in \sectionname{}~\ref{unclear} therefore persists in the non-Cowling case \citep[][]{Takata2016a}.

\subsection{Basic asymptotic expression of $\Theta_{\rm g}$ in RGB stars}

By injecting \eqs{theta_G}{int centre}{delta}~in \eq{theta_g_2}, the asymptotic expression of $\Theta_{\rm g}$ in RGB stars can be rewritten (modulo $\pi$)
\algn{
\Theta_{\rm g}=  \frac{\sqrt{2}}{\sigma} \int_{0}^{r_2} \frac{N}{r} \dd r + \int_{\tilde{r}}^{r_2} \left(\mathcal{K}_r - \frac{\sqrt{2}}{\sigma} \frac{N}{r}\right)\dd r -\Psi  \; .
\label{Theta_g final}
}
We retrieve Eq.~(136) of T16 who obtained this in a different way, that is through matching the wave functions coming from the different regions (i.e., from the center, evanescent region, and surface). It is also worth mentioning that, under the Cowling approximation, a term with a value of $\pi/2$ needs to be added to \eq{Theta_g final} owing to the different behavior of the wave function near $r_1$ in the Cowling hypothesis and in the non-Cowling case (see the second paragraph of \sectionname{}~\ref{link RGB}). As seen in \figurename{}~\ref{prop_diag} and \eq{K_r}, this is actually related to the change in nature of the innermost turning point $r_1$, which is located where $\sigma=\mathcal{S}_1(r_1)$ in the non-Cowling case instead of $\sigma=N(r_1)$ in the Cowling approximation. This modification leads to a shift of half the radial wavelength in the wave function, whence an additional phase of $\pi/2$ in $\Theta_{\rm g}$. This point emphasizes the importance of going beyond the Cowling approximation when studying the gravity offset.

According to \eq{Theta_g final}, identifying the first term in the right-hand side as the main contribution to the leading order term of $\Theta_{\rm g}$, we see that the variation in $\varepsilon_{\rm g}$ on the RGB certainly results from structural changes in the upper part of the radiative zone and the surrounding evanescent region (i.e., between about $\tilde{r}$ and $r_2$) via both the wavenumber integral and the $\Psi$ function. To relate the value of $\varepsilon_{\rm g}$ to the internal properties, we thus need to model this region.

%
\section{Evolution of $\varepsilon_{\rm g}$ on the red giant branch}
\label{evolution on RGB}

In this section, we investigate the evolution of $\varepsilon_{\rm g}$ on the RGB. By modeling in a simple way the Brunt-Väisalä and Lamb frequencies between the hydrogen-burning shell and the base of the convective zone, we first obtain an analytical expression for $\varepsilon_{\rm g}$, which is subsequently used to interpet the observed variations in \smash{$\varepsilon_{\rm g}^{\rm obs}$} during evolution.

\begin{figure}
\centering
\includegraphics[scale=0.3,trim= 0cm 0cm 0cm 0cm, clip]{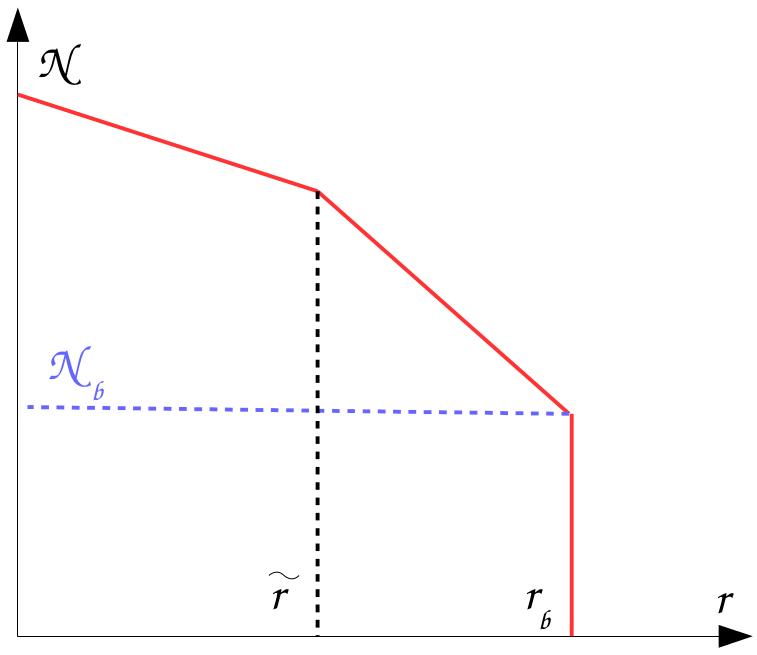} 
\caption{Schematic view of the modified Brunt-Väisälä frequency as a function of radius on a logarithmic scale, as assumed by the simple model presented in \sectionname{}~\ref{modeling}. The horizontal dashed line indicates the value $\mathcal{N}_{\rm b}$ just below the base of the convective zone at radius $r_{\rm b}$. The radius $\tilde{r}$ represents the location of the hydrogen-burning shell, in agreement with its definition given in \sectionname{}~\ref{link RGB}.} 
\label{scheme}
\end{figure}
%

\subsection{Simplified modeling of the upper radiative layers}
\label{modeling}

As shown by the propagation diagrams in \figurename{}~\ref{prop_diag}, the Brunt-Väisälä frequency varies at first approximation as a power law of radius between the hydrogen-burning shell and the base of the convective zone. This behavior is mainly due to the high density contrast existing between the helium core and the convective envelope on the RGB. For young red giant stars (e.g., model with $ \nu_{\rm max} \approx 200 ~\mu $Hz in \figurename{}~\ref{prop_diag}), the turning point $r_2$, where $\mathcal{N}(r_2) = 2 \pi \nu\sim 2\pi\nu_{\rm max}$ is located in the radiative region and depends on the mode frequency since \eq{drds} is non-null here. When stars evolve on the RGB, $\nu_{\rm max}$ decreases because of the envelope expansion. Indeed, $\nu_{\rm max}$ is about proportional to the acoustic cut-off frequency at the surface and its value is ruled by the scaling law \citep[e.g.,][]{Kjeldsen1995,Belkacem2013}
\algn{
\nu_{\rm max}\approx \nu_{{\rm max},\odot} \left( \frac{M}{M_\odot}\right)\left( \frac{R}{R_\odot}\right)^{-2}
\left( \frac{T_{\rm eff}}{T_{{\rm eff},\odot}}\right)^{-1/2} \; ,
\label{nu_max}
}
where $M$ is the stellar mass, $R$ is the stellar radius, and \smash{$\nu_{{\rm max},\odot} = 3104~\mu$Hz} is the solar reference value as prescribed by \cite{Mosser2013b}. Hence, since the increase in $R$ is predominant on the RGB, $\nu_{\rm max}$ decreases during evolution.
In contrast, the (usual and modified) Brunt-Väisälä frequency increases owing to the core contraction. As a result, the value of $\nu_{\rm max}$ becomes low enough at one point on the RGB for $r_2$ to be about equal to the radius of the base of the convective zone, which is denoted with $r_{\rm b}$. Since $r_2 \approx r_{\rm b}$ and $(\dd\ln \mathcal{N}/\dd \ln r)$ tends to minus infinity close to $r_{\rm b}$, \eq{drds} shows that the turning point $r_2$ then becomes frequency independent around $\nu_{\rm max}$ for the rest of the evolution on the RGB (e.g., model with $ \nu_{\rm max} \approx 60 ~\mu $Hz in \figurename{}~\ref{prop_diag}). Given that $\varepsilon_{\rm g}$ is sensitive to the properties of the
region in the vicinity of the turning points (see \sectionname{}~\ref{asymptotic epsilon_g}), its computation must account for this change of configuration.
In the following, we denote the value of the modified Brunt-Väisälä frequency just below the base of the convective region with $\mathcal{N}_{\rm b}$. We therefore distinguish two cases. First, case $a$: young stars on the beginning of the ascent of the RGB for which $r_2 < r_{\rm b}$ and $2 \pi \nu_{\rm max} > \mathcal{N}_{\rm b}$; and second, case $b$: more evolved stars with $r_2 \approx r_{\rm b}$ and \smash{$2 \pi \nu_{\rm max} \lesssim \mathcal{N}_{\rm b}$.}

Moreover, to express $\varepsilon_{\rm g}$, we use several simplifying assumptions. First, the (modified) Brunt-Väisälä frequency is modeled as a decreasing power law of radius with an index $\beta$ for \smash{$\tilde{r} \le r \le r_{\rm b}$,} that is,
\algn{
\mathcal{N}(r)=\mathcal{N}_{\rm b}\left(\frac{r_{\rm b}}{r} \right)^\beta \; .
\label{power law}
}
At the base of the convective zone, $\mathcal{N}$ varies so sharply close to $r_{\rm b}$ that it is assumed to discontinuously change from $\mathcal{N}_{\rm b}$ to about zero. A schematic view of this simplified configuration is shown in \figurename{}~\ref{scheme}. In case $a$ only, we note that
\algn{
\mathcal{N}_{\rm b} r_{\rm b}^\beta =\sigma r_2^\beta \; ,
\label{power law 2}
}
since $\mathcal{N}(r_2) = \sigma$ at the turning point. Second, for sake of simplicity, we also focus on the case of a thick evanescent region, so that $\sigma \ll \mathcal{S}_1$ everywhere in the buoyancy cavity and $\Psi \approx - 1/4$. Third and lastly, the relation \smash{$J \approx {\rm const.} \approx 2 \beta /3$}, which results from the high density contrast between the core and the envelope, is supposed to be met in a good approximation between the hydrogen-burning shell and the base of the convective zone (e.g., see \sectionname{}~A.3.1 of T16). All these hypotheses are checked a posteriori in \sectionname{}~\ref{discussion assumption} not to modify the final conclusions.

\subsection{Asymptotic expressions of $\Delta \Pi_1$ and $\varepsilon_{\rm g}$}
\label{as eps}

Under the previous assumptions, we have to compute the integrals between $\tilde{r}$ and $r_2$ in \eq{Theta_g final} to completely express $\Theta_{\rm g}$ and then identify $\varepsilon_{\rm g}$ and $\Delta \Pi_1$ based on \eqss{Theta_g formal}{local epsilon_g}.

\subsubsection{Computations in case $a$ $(r_2  < r_{\rm b}, ~\sigma > \mathcal{N}_{\rm b})$}
\label{case $a$}

The case $a$ has already been considered by \cite{Takata2016a}. Nevertheless, most of the analytical developments in this case are useful to treat case $b$. For sake of clarity and convenience, we thus propose in the following to detail the computation, considering the hypothesis of a thick evanescent region only. 

The first step consists in expressing the frequency dependence of $r_2$ in the upper bound of the first integral that appears in the right-hand side of \eq{Theta_g final}, denoted with $\mathcal{I}_1$ in the following. This latter can be rewritten as 
\algn{
\mathcal{I}_1 &=\int_{0}^{r_{\rm b}}\frac{\sqrt{2}}{ \sigma} \frac{N}{r} \dd r -\int_{r_2}^{r_{\rm b}} J\frac{\sqrt{2}}{ \sigma}\frac{\mathcal{N}}{r} \dd r \label{I_20} \; .
}
In case $a$, it results to
\algn{\mathcal{I}_1^a&= \int_{0}^{r_{\rm b}} \frac{\sqrt{2}}{ \sigma}\frac{N}{r} \dd r - \frac{\sqrt{2} J}{\beta} \left(1-\frac{\mathcal{N}_{\rm b}}{\sigma}\right) \; ,
\label{I_2}
}
where the second integral in \eq{I_20} is computed using the change of variable $y=\mathcal{N}$ with \smash{$y(r_2)=\sigma$} and $y(r_{\rm b})=\mathcal{N}_{\rm b}$.

The second step is to consider the second integral in the right-hand side of \eq{Theta_g final}, denoted with $\mathcal{I}_2$ in the following. Assuming $\sigma \ll \mathcal{S}_1$ in \eq{K_r}, this integral can be rewritten
\algn{
\mathcal{I}_2\approx \int_{\tilde{r}}^{r_2} \sqrt{2} J \left[ \left(\frac{\mathcal{N}^2}{\sigma^2}-1 \right)^{1/2}-\frac{\mathcal{N}}{\sigma}\right] \frac{\dd r}{r} \; .
\label{I_10}
}
Then, using the change of variable $x=\sigma/\mathcal{N}$ and the fact that $x(r_2)=1$, we obtain in case $a$ \citep[e.g.,][]{Abramowitz1972}
\algn{
\mathcal{I}_2^a&\approx \frac{\sqrt{2} J}{\beta} \int_{x_\star}^{1} \left(\sqrt{1-x^2}-1 \right) \frac{\dd x}{x^2} \\
&= \frac{\sqrt{2} J}{\beta} \left[ 1-\frac{\pi}{2}+O\left(x_\star\right)\right] \; ,
\label{I_1}
}
where $x_\star=\sigma/\mathcal{N}(\tilde{r}) \ll1$ in the asymptotic limit and is neglected in the following.

The final step consists in injecting \eqs{I_2}{I_1}~in \eq{Theta_g final}, so that the asymptotic expression of $\Theta_{\rm g}$ finally reads in case $a$
\algn{
\Theta_{\rm g}^a\approx\frac{\sqrt{2}}{ \sigma} \left(\int_{0}^{r_{\rm b}} \frac{N}{r} \dd r + \frac{N_{\rm b}}{\beta} \right) - \frac{\sqrt{2} J \pi}{2 \beta} - \Psi \; ,
\label{Theta_g YR}
}
where $N_{\rm b} = J \mathcal{N}_{\rm b} $ is the value of the usual Brunt-Väisälä frequency just below $r_{\rm b}$. According to \eq{Theta_g formal}, the $F$ function is equal in case $a$ to 
\algn{
F^a(\nu)\approx \frac{\sqrt{2}}{ 2\pi} \left(\int_{0}^{r_{\rm b}} \frac{N}{r} \dd r + \frac{N_{\rm b}}{\beta} \right) - \left( \frac{\sqrt{2} J \pi }{2 \beta} +  \Psi  \right) \nu\; .
\label{F a}
}
In the hypothesis of a thick evanescent region, $\Psi$ is about constant and close to $ -\pi/4$ (see \sectionname{}~\ref{link RGB}). The determination of the local asymptotic period spacing and gravity offset around $\nu_{\rm max}$ is thus obtained from \eqs{local Delta Pi}{local epsilon_g}, that is,
\algn{
\Delta \Pi_1^a= \frac{2 \pi^2}{ \sqrt{2}} \left(\int_{0}^{r_{\rm b}} \frac{N}{r} \dd r + \frac{2 \mathcal{N}_{\rm b}}{3} \right)^{-1} \; ,
\label{Delta Pi YR}
}
and 
\algn{
\varepsilon_{\rm g}^a= \frac{\sqrt{2} J}{2\beta} + \frac{\Psi}{\pi} \approx \frac{\sqrt{2}}{3} - \frac{1}{4} \approx 0.22\; ,
\label{epsilon_g a}
}
where we used the relation $\beta\approx3 J/2$ and $N_{\rm b} \approx (2 \beta /3) \mathcal{N}_{\rm b}$. We retrieve the result found by T16 in the hypothesis of a thick evanescent region.
We note that the terms in brackets in the right-hand side of \eq{Delta Pi YR} would be equal to the integral between $0$ and $+\infty$ of the Brunt-Väisälä frequency if this latter hypothetically varied following \eq{power law} from $\tilde{r}$ to $+\infty$.

Therefore, the model predicts that $\varepsilon_{\rm g}^a$ is independent of $\nu_{\rm max}$ so that the gravity offset remains constant during evolution for the young red giant stars in case $a$. Actually, according to \eq{F a}, the local first-order analysis around $\nu_{\rm max}$ appears to be a very good approximation in case $a$ since the remainder in \eq{gamma} originates only from the residual in \eq{I_1} that is on the
order of $x_\star$. Indeed, we can check that $x_\star \sim 0.01$ in typical red giant stars so that the remainder is clearly negligible (e.g., see \figurename{}~\ref{prop_diag}). This result validates the assumptions made in the measurement of \smash{$\Delta \Pi_1^{\rm obs}$} and \smash{$\varepsilon_{\rm g}^{\rm obs}$} for stars in case $a$ (see \sectionname{}~\ref{observational definition}). We conclude that the local asymptotic values of $\Delta \Pi_1$ and $\varepsilon_{\rm g}$ are representative of both observed quantities, respectively; thus their comparison is relevant. Actually, the same conclusions can be reached in the hypothesis of a very thin evanescent region, as discussed in \sectionname{}~\ref{discussion assumption}.

\subsubsection{Computations in case $b$ $(r_2 \approx r_{\rm b}, ~\sigma < \mathcal{N}_{\rm b})$}
\label{case $b$}
\begin{figure}
\centering
\includegraphics[scale=0.35,trim= 0.5cm 0cm 2cm 0cm, clip]{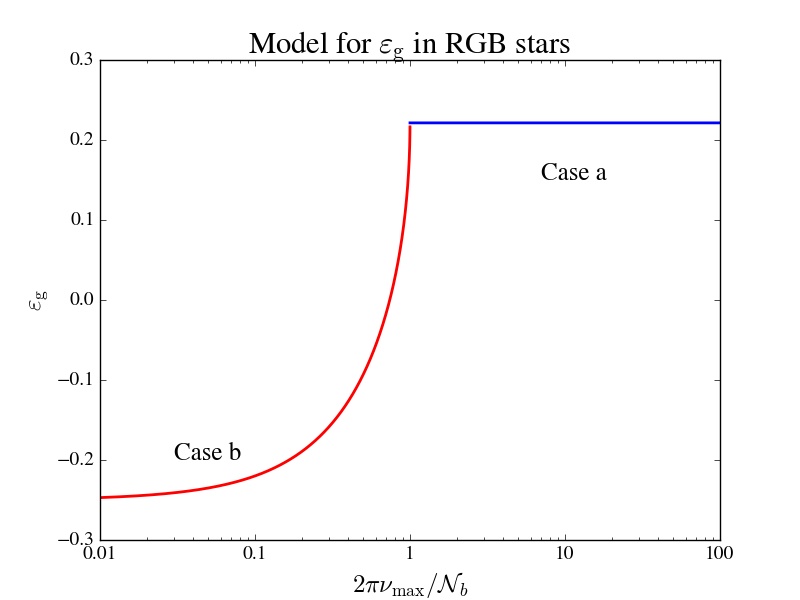} 
\caption{Local asymptotic value of the gravity offset according to the model considered in \sectionname{}~\ref{modeling}. It was computed via \eq{epsilon_g a} in case $a$ (blue) and \eq{epsilon_g b} in case $b$ (red). The cases $a$ and $b$ are defined in \sectionname{}~\ref{modeling}. It is shown as a function of the ratio $2 \pi \nu_{\rm max}/\mathcal{N}_{\rm b}$, where $\mathcal{N}_{\rm b}$ is the value of the (modified) Brunt-Väisälä frequency just below the base of the convective zone. Stellar evolution goes from the right to the left.} 
\label{offset}
\end{figure}

For stars in case $b$, the upper bound of the first integral in \eq{Theta_g final} is frequency independent since $r_2$ must be replaced by $r_{\rm b}$, so that this integral is merely equal in case $b$ to
\algn{
\mathcal{I}_1^b &=\int_{0}^{r_{\rm b}} \frac{\sqrt{2}}{ \sigma} \frac{N}{r} \dd r  \; .
\label{I_2 b}
}
To go further, the computation of the second integral in \eq{Theta_g final} is similar to \eqss{I_10}{I_1}~in case $a$, except that the upper bound of the integral is now frequency dependent and equal to \smash{$x_{\rm b} =2 \pi \nu/\mathcal{N}_{\rm b}$.} As a result, we obtain in case $b$ \citep[e.g.,][]{Abramowitz1972}
\algn{
\mathcal{I}_2^b(x_{\rm b}) \approx  \frac{2 \sqrt{2} }{3} \left(\frac{1-\sqrt{1-x_{\rm b}^2}}{x_{\rm b}} -\arcsin(x_{\rm b})\right) \; ,
\label{I_1 b}
}
where we again used the relation $\beta \approx 3 J/2$ and the fact that \smash{$x_\star \ll 1$.}
Injecting \eqs{I_2 b}{I_1 b}~in \eq{Theta_g final}, we get
\algn{
\Theta_{\rm g}^b\approx \frac{\sqrt{2}}{ \sigma} \int_{0}^{r_{\rm b}} \frac{N}{r} \dd r +\mathcal{I}_2^b\left(\frac{\sigma}{\mathcal{N}_{\rm b}}\right)-\Psi\; .
\label{Theta_g ER}
}
Proceeding in a similar way as in \sectionname{}~\ref{case $a$},
the $F$ function now reads in case $b$, according to \eq{Theta_g formal},
\algn{
F^b(\nu) \approx \frac{\sqrt{2}}{ 2\pi} \left(\int_{0}^{r_{\rm b}} \frac{N}{r} \dd r \right) + \frac{\mathcal{N}_{\rm b}}{2\pi}  \left[ \mathcal{I}_2^b(x_{\rm b})-  \Psi  \right] x_{\rm b} \; .
\label{F b}
}
In the hypothesis of a thick evanescent region, $\Psi$ is a constant close to $-\pi/4$ and the derivative of $F$ with respect to frequency is thus equal to
\algn{
F^{b\prime}(\nu) \approx - \frac{2 \sqrt{2} }{3} \arcsin(x_{\rm b}) - \Psi\; .
\label{F p b}
}
As a final step, the local asymptotic period pacing and gravity offset can be computed using \eqs{local Delta Pi}{local epsilon_g}{F b}{F p b}, that is,
\algn{
\Delta \Pi_1^b = \frac{2 \pi^2}{ \sqrt{2}}\left[\int_{0}^{r_{\rm b}} \frac{N}{r} \dd r + \frac{2 \mathcal{N}_{\rm b}}{3} \left(1-\sqrt{1-\left(\frac{2\pi \nu_{\rm max}}{\mathcal{N}_{\rm b}}\right)^2}~\right)\right]^{-1}\; ,
\label{Delta Pi ER}
}
and
\algn{
\varepsilon_{\rm g}^b &= \frac{2 \sqrt{2} }{3 \pi} \arcsin\left(\frac{2\pi \nu_{\rm max}}{\mathcal{N}_{\rm b}} \right) + \frac{\Psi}{\pi} \approx \frac{2 \sqrt{2} }{3 \pi} \arcsin\left(\frac{2\pi \nu_{\rm max}}{\mathcal{N}_{\rm b}} \right)-\frac{1}{4} \; .
\label{epsilon_g b}
}
Unlike in case $a$, the local asymptotic gravity offset depends on $\nu_{\rm max}$ in case $b$ and therefore must vary during evolution. In this case, the remainder around $\nu_{\rm max}$ defined in \eq{gamma} is non-null and is equal using \eqs{Theta_g formal}{Taylor F}{gamma}{F b}{F p b}~to
\algn{
\gamma(\nu;\nu_{\rm max})&\approx\frac{1}{\pi} \left[\mathcal{I}_2^b\left(x_{\rm b}\right)-\mathcal{I}_2^b\left(x_{{\rm b,max}}\right) \right]\nonumber \\
 &+\frac{2 \sqrt{2} }{3 \pi}\left(\sqrt{1-x_{{\rm b,max}}^2}~-1\right)\left(\frac{1}{x_{\rm b}}-\frac{1}{x_{{\rm b,max}}}\right) \; ,
 \label{gamma b}
}
where $x_{{\rm b,max}}=2\pi \nu_{\rm max}/\mathcal{N}_{\rm b}$. We show in \sectionname{}~\ref{validity local} that the remainder is not higher than about $\sigma_\varepsilon$ over the observed frequency range of typical red giant stars in case $b$ and thus can be neglected. As a consequence, the definitions of the local asymptotic period spacing and gravity offset given in \eqs{local Delta Pi}{local epsilon_g}~are representative of the observed values and the comparison between both remains relevant in case $b$.

\subsection{Variation of $\varepsilon_{\rm g}$ with evolution}

\subsubsection{Asymptotic predictions of the model}
\label{prediction}

The variation of the gravity offset during the evolution on the RGB can now be discussed in the framework of the model developed in the previous sections. From \eqs{epsilon_g a}{epsilon_g b}, it is possible to compute $\varepsilon_{\rm g}$ as a function of the parameter $2\pi \nu_{\rm max}/\mathcal{N}_{\rm b}$. This ratio monotically decreases with time and thus represents a proxy of the evolutionary stage on the RGB (see \appendixname{}~\ref{Nb evol}). As expected from stellar models of typical observed RGB stars, the value of this parameter is considered between 0.01 and 100. The result is shown in \figurename{}~\ref{offset}.

For red giant stars such as $2 \pi \nu_{\rm max} \ge \mathcal{N}_{\rm b}$ (case $a$), the model predicts that the value of $\varepsilon_{\rm g}$ remains constant and about equal to 0.22. According to \eq{epsilon_g a}, this peculiar value results from the power-law behavior of the Brunt-Vaïsälä frequency and from the relation between $J$ and $\beta$, which are a direct consequence of the high contrast of density between the helium core and the base of the convective envelope in these stars. Once $2\pi \nu_{\rm max} /\mathcal{N}_{\rm b}=1$, the value of $\varepsilon_{\rm g}$ rapidly drops to reach negative values. This sharp decrease is the consequence of the kink of the Brunt-Väisälä frequency profile near the base of the convective zone. Indeed, while the turning point $r_2$ depends on the oscillation frequency when $2\pi \nu_{\rm max}  > \mathcal{N}_{\rm b} $, the change in the behavior of $\mathcal{N}$ near $r_2$ in evolved red giant stars such as $2 \pi \nu_{\rm max} < \mathcal{N}_{\rm b}$ makes $r_2$ frequency independent. As a consequence, the term $\sqrt{2} J/\beta \approx 2 \sqrt{2} /3$ in \eq{I_2}, which contributes to the value of $\varepsilon_{\rm g}$ in case $a$, disappears in \eq{I_2 b}. In parallel, $\mathcal{K}_r$ becomes discontinuous at $r_2$ in case $b$ and thus does not tend to zero as $r$ tends to $r_2$. The frequency independent integral in \eq{I_1} for the case $a$ is thus replaced by the frequency dependent integral in \eq{I_1 b} for the case $b$. As a result, $\varepsilon_{\rm g}$ in case $b$ depends on $\nu_{\rm max}$. Moreover, its derivative with respect to $\nu_{\rm max}$ turns out to be discontinuous, going from 0 to $+\infty$ at the transition between both cases (in the sense of the evolution), as seen in \figurename{}~\ref{offset}. Using standard stellar models of $1$, $1.2$, $1.4$, $1.6$, $1.8,$ and $2M_\odot$ computed by means of the sellar evolution code CESTAM, the input physics of which is detailed in \sectionname{}~\ref{unclear}, we estimated based on the scaling relation in \eq{nu_max} that this transition must happen for values of \smash{$\nu_{\rm max, t}\approx 110$, 95, 80, 70, 60, and 50$~\mu$Hz,} respectively, that is, just before the luminosity bump \citep[e.g,][]{JCD2015}. We note that the higher the mass, the lower the value of $\nu_{\rm max, t}$. Interestingly, \cite{Khan2018} found a similar trend in the observed value of $\nu_{\rm max}$ at the luminosity bump.

As $2 \pi \nu_{\rm max}/\mathcal{N}_{\rm b}$ keeps decreasing with evolution after the transition, the model predicts that $\varepsilon_{\rm g}$ smoothly decreases. Indeed, the integral in \eq{I_1 b}, which actually represents the error made by computing the wavenumber integral in the upper part of the buoyancy cavity using the approximate expression for the radial wavenumber given in \eq{K_r approx}, tends to zero as the oscillation frequency decreases. In the case of very evolved stars for which \smash{$2 \pi \nu_{\rm max} \ll \mathcal{N}_{\rm b}$}, the value of the gravity offset is then expected to approach $-0.25$, that is, the limit value of the phase $\Psi$ as the evanescent region becomes very thick.

\subsubsection{Observed values}

\cite{Mosser2018} measured the value of the gravity offset in 196 RGB stars observed by the {\it Kepler} satellite, that is, with values of the period spacing lower than $150$~s and in a mass range between about $1M_\odot$ and $2M_\odot$ \citep[e.g.,][]{Mosser2014}. The median uncertainty on $\varepsilon_{\rm g}^{\rm obs}$ for this sample of stars is about $\sigma_{\varepsilon}\sim 0.08$, which turns out to be large for a parameter smaller than unity. Given the intrinsic difficulty of measuring this parameter, we thus consider only the stars with error bars on $\varepsilon_{\rm g}^{\rm obs}$ below $2\sigma_{\varepsilon} \sim 0.16$. The resulting selected sample consists in 187 red giant stars. We checked that the rejected stars do not belong to a peculiar subset regarding the mass and evolutionary state and that the observed values of the coupling factor, period spacing, and large separation in these stars follow the same trend as the whole sample. Further analyses of the power spectra of the rejected stars are nevertheless needed to fully confirm that such large uncertainties originate from errors in the fits rather than structural peculiarities, which is out of scope of this paper.

\subsubsection{Comparison: Model versus observations}
\label{comparison}

\begin{figure*}
\centering
\includegraphics[scale=0.5,trim= 0.5cm 0cm 2cm 1cm, clip]{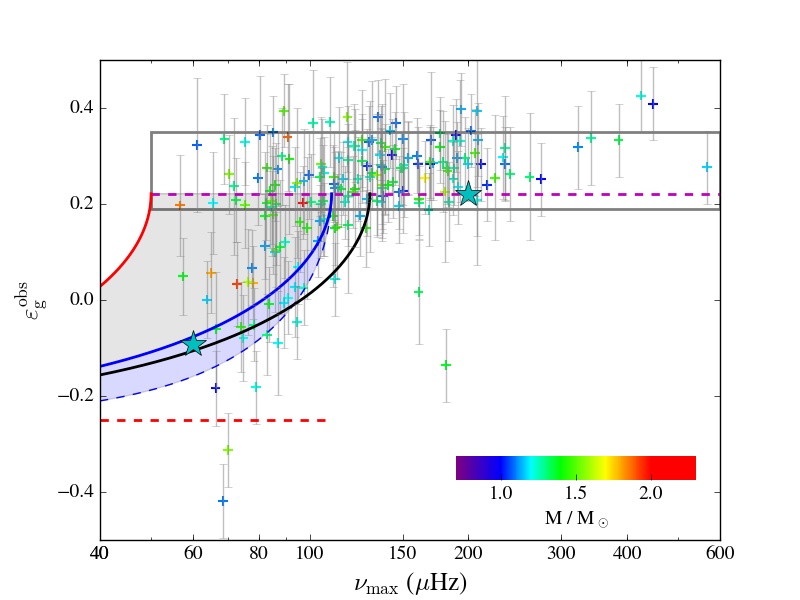} 
\caption{Observed values of the gravity offset adapted from \cite{Mosser2018} as a function of $\nu_{\rm max}$ in RGB stars ($+$ symbols). The color code indicates the seismic mass computed following \cite{Mosser2013b}. The error bars are also represented for each data point. The horizontal magenta dashed line corresponds to the asymptotic value expected from \eq{epsilon_g a} for RGB stars in case $a$ and in the low coupling hypothesis (see \sectionname{}~\ref{modeling} for the definition of the cases $a$ and $b$). This value is included in the dark gray rectangle that corresponds to the values predicted by \cite{Takata2016a}, considering typical structures of stars in the early RGB phase (see \sectionname{}~\ref{discussion assumption}). The clear gray domain represents the asymptotic prediction that is expected from \eq{epsilon_g b} for RGB stars in case $b$. Guided by stellar models, the transition is assumed to occur between \smash{$\nu_{\rm max} \approx 50~\mu$Hz} (red solid line) and \smash{$\nu_{\rm max} \approx 110~\mu$Hz} (blue solid line), depending on stellar mass (see main text). The prediction of the analytical model is plotted as a function of $\nu_{\rm max}$ by assuming that the parameter $\mathcal{N}_{\rm b}$ in \eq{epsilon_g b} is constant. The black solid line shows the prediction for a transition at $\nu_{\rm max} = 130~\mu$Hz. In contrast, the clear blue domain corresponds the prediction for a value at the transition of $\nu_{\rm max,t}=110~\mu$Hz, assuming that $\mathcal{N}_{\rm b} \propto \nu_{\rm max}^x$ with $x$ between 0 and 1 (see \sectionname{}~\ref{comparison}). The result with $x=1$ is represented by the blue dashed line. The horizontal red dashed line indicates the lower limit predicted by the model for stars in case $b$. As an illustration, the predicted values of $\varepsilon_{\rm g}$ for the two 1.2$M_\odot$ stellar models considered in \figurename{}~\ref{prop_diag} are represented by the cyan stars.}
\label{obs}
\end{figure*}

The values of the gravity offset in the 187 selected RGB stars are reported in \figurename{}~\ref{obs} as a function of $\nu_{\rm max}$. As shown in \figurename{}~\ref{obs}, the authors noticed an accumulation of values in the range $[0.2,0.35]$ for $\nu_{\rm max} \gtrsim 110~\mu$Hz. In contrast, for $\nu_{\rm max} \lesssim 110~\mu$Hz, \figurename{}~\ref{obs} shows that \smash{$\varepsilon_{\rm g}^{\rm obs}$} rapidly decreases in average by an amount of about $0.2$ and that its value can even becomes negative for several stars.

To compare the observations with the theoretical predictions of the analytical model, we need to make assumptions on the value of $\nu_{\rm max}$ at the transition between cases $a$ and $b$, denoted with $\nu_{\rm max,t}$ (i.e., for which $2 \pi \nu_{\rm max}=\mathcal{N}_{\rm b}$), and on the evolution of the ratio $2\pi \nu_{\rm max}/\mathcal{N}_{\rm b}$ for stars in case $b$. First, we assume that $\nu_{\rm max,t}$ takes values between $50~\mu$Hz and $110~\mu$Hz in the considered mass range, as estimated through standard stellar models of typical red giant stars (see \sectionname{}~\ref{unclear} for details on the input physics). Second, as suggested by the 1.2$M_\odot$ stellar models in \figurename{}~\ref{prop_diag}, $\mathcal{N}_{\rm b}$ generally increases between the beginning of the RGB and the luminosity bump while $\nu_{\rm max}$ decreases. Indeed, a rough estimate of $\mathcal{N}_{\rm b}$ just below the base of the convective zone is provided according to the definition of $\mathcal{N}$ with $J(r_{\rm b})\sim 1$ by 
\algn{
\mathcal{N}_{\rm b}\sim \frac{G m(r_{\rm b})}{r_{\rm b}^3} \; ,
\label{Nb}
}
where $G$ is the gravitational constant and $m(r)$ represents the mass inside the shell of radius $r$. Using stellar models in the considered mass range, we checked that \eq{Nb} is a good proxy of the actual value of the Brunt-Väisälä frequency just below the base of the convective envelope since it follows the same trend during evolution. At the beginning of the ascent of the RGB, the inner convective boundary deepens inside the star until a point just before the luminosity bump where it starts receding. This leads to the well-known first convective dredge-up of the chemical elements. In the first part of the RGB, $\mathcal{N}_{\rm b}$ increases because of the predominant effect of the decrease in $r_{\rm b}$. In the subsequent short recession phase after the convective boundary has attained its minimum and before the luminosity bump, $\mathcal{N}_{\rm b}$ can be regarded as about constant. In parallel, the expansion of the envelope is so rapid at this evolutionary state that $\nu_{\rm max}$ decreases more rapidly than $\mathcal{N}_{\rm b}$ increases. This trend is observed in all the considered mass range (see \appendixname{}~\ref{Nb evol} for more details). In order to model this in a simple way, 
we assume in the following that $\mathcal{N}_{\rm b}$ decreases as $\nu_{\rm max}^{-x}$ with $x$ a real between zero and unity. Under this hypothesis, we have
\algn{
\frac{2\pi \nu_{\rm max} }{\mathcal{N}_{\rm b}} =\left( \frac{\nu_{\rm max}}{\nu_{\rm max,t}}\right)^{1+x} \; .
\label{ratio assumption}
}
As a first step, the decrease in the ratio can be supposed to be mainly due to the decrease in $\nu_{\rm max}$, and therefore $\mathcal{N}_{\rm b}$ can be considered as a constant (i.e., $x=0$).
The resulting predicted value of $\varepsilon_{\rm g}$ in case $b$, for $\nu_{\rm max,t} = 50-110~\mu$Hz, are represented by the light gray domain in \figurename{}~\ref{obs}. As a test, the corresponding prediction of the model for a slightly higher value of $\nu_{\rm max,t}$ at 130~$\mu$Hz, which corresponds to a transition occurring at an earlier evolutionary state in the considered mass range, is represented by the black solid line. In the opposite limiting case, $\mathcal{N}_{\rm b}$ can be supposed to increase as rapidly as $\nu_{\rm max}$ decreases (i.e., $x=1$). The result is indicated in the same figure by the blue dashed line for $\nu_{\rm max,t}=110~\mu$Hz. The clear blue domain represents the analytical predictions for $x$ between zero and unity at $\nu_{\rm max,t}=110~\mu$Hz. We see that the higher the value of $x$, the sharper the decrease in $\varepsilon_{\rm g}$ as $\nu_{\rm max}$ decreases.

Figure~\ref{obs} shows that observations and theory are in a noteworthy agreement. On the one hand, the observed values in stars with $\nu_{\rm max} \gtrsim 110~\mu$Hz and their error bars are compatible with the constant asympotic value of 0.22 predicted by the analytical model. On the other hand, the global drop in \smash{$\varepsilon_{\rm g}^{\rm obs}$} for $\nu_{\rm max} \lesssim 110~\mu$Hz is also well reproduced by the theoretical expectations. Indeed, the rapid decrease in \smash{$\varepsilon_{\rm g}^{\rm obs}$} occurs about at the same evolutionary stage as predicted through stellar models, that is, for $\nu_{\rm max} ~\sim 50-110~\mu$Hz in the considered mass range. Moreover, the mean amplitude of the drop in $\varepsilon_{\rm g}^{\rm obs}$ reasonably agrees with the predictions of the analytical model. In the framework of this model, we thus conclude that the stars for which the value of $\varepsilon_{\rm g}^{\rm obs}$ is slightly lower than about $0.22$ correspond to stars that have just spent the transition between cases $a$ and $b$ (i.e., such as $2 \pi \nu_{\rm max} /\mathcal{N}_{\rm b} \sim 1$) and for which the gravity offset starts rapidly decreasing. Furthermore, it is also worth mentioning that less massive stars visually seem to undergo the transition and take negative values before the more massive stars, as predicted through stellar models (see \sectionname{}~\ref{prediction} and \appendixname{}~\ref{Nb evol}). However, given the small amount of data for the less and the more massive stars of the sample at the transition compared to other observed stars, a statistical test of this trend currently appears to be difficult. In later stages, for $\nu_{\rm max} \lesssim 50~ \mu$Hz, we expect to observe values between about $-0.1$ and $-0.25$. Again, because of the difficulty to observe gravity-dominated mixed modes in these stars \cite[e.g.,][]{Grosjean2014}, there are few data available to statistically test this prediction.

For the frequency range $\nu_{\rm max,t} =50-110~\mu$Hz, \figurename{}~\ref{obs} shows that accounting for the increase in $\mathcal{N}_{\rm b}$ along with evolution after the transition extends the predicted domain toward lower values of $\varepsilon_ {\rm g}$ (i.e., see the clear blue domain in \figurename{}~\ref{obs}). As an illustration, we can consider the case of the $1.2M_\odot$ red giant model with $\nu_{\rm max}\approx 60~\mu$Hz in \figurename{}~\ref{prop_diag}. For this stellar model, we estimated $2\pi \nu_{\rm max} / \mathcal{N}_{\rm b} \approx 0.5$ so that the value of $\varepsilon_{\rm g}$ predicted by \eq{epsilon_g b} is close to $-0.1$, as reported in \figurename{}~\ref{obs}. As said in \sectionname{}~\ref{prediction}, we also estimated from its evolutionary sequence that $\nu_{\rm max,t} \approx 95~\mu$Hz at the transition. If $\mathcal{N}_{\rm b}$ was constant, we would have, in contrast, $2\pi \nu_{\rm max} / \mathcal{N}_{\rm b}= \nu_{\rm max} / \nu_{\rm max,t} \approx 0.63$ from \eq{ratio assumption} and $\varepsilon_{\rm g}\approx -0.04$ according to \eq{epsilon_g b}. Therefore, accounting for the increase in $\mathcal{N}_{\rm b}$ is necessary for a value of $\varepsilon_{\rm g}=-0.1$ to be compatible with the value of $\nu_{\rm max,t} \approx 95~\mu$Hz at the transition in this standard stellar model. Nevertheless, \figurename{}~\ref{obs} shows that slightly higher values of $\nu_{\rm max,t}$ can also lead on their own to lower values of $\varepsilon_{\rm g}$ at a given value of $\nu_{\rm max}$ in case $b$ (i.e., see the black solid line in \figurename{}~\ref{obs}). Therefore, a value of $\varepsilon_{\rm g}=-0.1$ at $\nu_{\rm max}\approx 60~\mu$Hz can be explained not only by an increase in $\mathcal{N}_{\rm b}$ during evolution for $\nu_{\rm max,t} \approx 95~\mu$Hz, as in the $1.2M_\odot$ red giant model considered in \figurename{}~\ref{prop_diag}, but also by a slightly higher value of $\nu_{\rm max,t}$ between $95~\mu$Hz and $130~\mu$Hz at a constant value of $\mathcal{N}_{\rm b}$, or by the simultaneous effect of both. While the progressive increase in $\mathcal{N}_{\rm b}$ during evolution results from the inward migration of the base of the convective zone at this evolutionary stage, the increase in the value of $\nu_{\rm max,t}$ can physically result from the presence of overshooting at the inner convective boundary. This point is discussed in more detail in \sectionname{}~\ref{overshooting}. We emphasize that both effects tend to improve the agreement with observations in case $b$, compared to the predictions obtained with the range of values for $\nu_{\rm max,t}$ that are estimated through standard stellar models without overshooting and considering $\mathcal{N}_{\rm b}$ as a constant during evolution.

Nevertheless, we notice four stars that exhibit discrepancies compared to the observed general trend and theoretical predictions. 
First, for $\nu_{\rm max} \sim 70~\mu$Hz, the stars KIC 3749487 and KIC 10387370 have \smash{$\varepsilon_{\rm g}^{\rm obs} \approx -0.3\pm 0.08$} and \smash{$\varepsilon_{\rm g}^{\rm obs} \approx -0.4\pm 0.08$}, respectively. These observations turn out to be well lower than the predicted range of values. Actually, the error bars in KIC 3749487 star are compatible with the lower limit of $-0.25$ expected from the model. In this star, the abnormal value of $\varepsilon_{\rm g}$ could therefore be explained within the framework of the model by a much more rapid increase in $\mathcal{N}_{\rm b}$ with evolution and/or a higher value of $\nu_{\rm max,t}$ at the transition. In contrast, the value measured in KIC 10387370 is too low compared to the predicted lower limit and thus is incompatible with the model. Second, for $150 ~\mu$Hz$\lesssim\nu_{\rm max}\lesssim 200~\mu$Hz, the stars KIC 2584478 and KIC 12008916 exhibit values of \smash{$\varepsilon_{\rm g}^{\rm obs}$} that are well below the values measured in dozens of other stars at the same evolutionary stage with similar masses but in agreement with the predictions of the model. We checked that the seismic properties of these four stars (e.g., period spacing, coupling factor, core rotation rate; see \tablename{}~\ref{table2}) follow the mean trend of the whole considered RGB sample, except the gravity offset. Rather than different structures, errors in the observed fits or underestimated uncertainties are thus certainly the most plausible explanation for these outliers. The four outliers therefore do not statistically hinder the global agreement between observations and theory. More detailed analyses of these four stars are nonetheless needed to attest the reasons for their deviations from the global trend, which is beyond the scope of this work.

In conclusion, the general trend observed in the variation of $\varepsilon_{\rm g}^{\rm obs}$ with evolution can be globally explained through the asymptotic expressions and the model that we developed. Red giant stars with $\nu_{\rm max} \gtrsim 110~\mu$Hz exhibit values of the gravity offset about constant and close to $0.22$. This mean value is the consequence of the high contrast of density between the core and the envelope. For red giant stars with $\nu_{\rm max} \sim 50-110~\mu$Hz, the rapid decrease observed in the gravity offset by a mean value of about $0.2$ occurs when $\nu_{\rm max}$ becomes so low that the outermost turning point $r_2$ associated with the buoyancy cavity becomes and remains about equal to the radius of the base of the convective zone, $r_{\rm b}$, during the subsequent evolution on the RGB. This is the consequence of the kink in the Brunt-Väisälä frequency profile at the interface between the radiative and convective regions. This clear signature in $\varepsilon_{\rm g}^{\rm obs}$ happens just before the luminosity bump. Furthermore, accounting for the effects of the inward migration of the base of the convective or of overshooting in this region tends to improve the agreement between the model and observations in the more evolved red giant stars with $\nu_{\rm max}\lesssim 110 \mu$Hz.

To finish, we emphasize that the agreement between observations and theory is remarkable given the difficulty to measure \smash{$\varepsilon_{\rm g}^{\rm obs}$}. Indeed, since the number of radial nodes in the buoyancy cavity is very large and typically close to $n_{\rm g}\sim 100$ in red giant stars, a difference as small as 0.5\%
between \smash{$\Delta \Pi_1^{\rm obs}$} and \smash{$\Delta \Pi_1$} leads to a difference of values of 0.5 between \smash{$\varepsilon_{\rm g}^{\rm obs}$} and \smash{$\varepsilon_{\rm g}$}. The present result therefore confirms the robustness of both the measurement method and the validity of the asymptotic analyses in red giant stars.

%
\section{Discussion}
\label{discussion} 

In this last section, we discuss the results and main assumptions used in this work.

\subsection{Validity of the local asymptotic definitions in RGB stars}
\label{validity local}

In this work, we considered local definitions around $\nu_{\rm max}$ for $\varepsilon_{\rm g}$ and $\Delta \Pi_1$ (see \sectionname{}~\ref{asymptotic definition}). The comparison between the local asymptotic expressions and observed values is relevant if both quantities are equivalent over all the observed frequency range around $\nu_{\rm max}$, which can be checked a posteriori.

For stars in case $a$, we showed in \sectionname{}~\ref{case $a$} that the local asymptotic expressions are representative of the observed values. In the framework of the considered model, the equivalence between both actually turned out to be easy to confirm. For stars in case $b$, the answer to this question is less straightforward. In order to check the validity of comparing local asymptotic expressions to observations in case $b$, we can proceed as follows. We still assume in a good approximation that the observed frequencies belong to a $6\Delta \nu$-wide frequency range centered around $\nu_{\rm max}$ as expected from theoretical estimates and observations in typical RGB stars \citep[e.g.,][]{Grosjean2014,Mosser2018}. Using the $\Delta \nu$-$\nu_{\rm max}$ scaling law for RGB stars, that is \citep[e.g.,][]{Peralta2018},
\algn{
\Delta \nu \approx 0.28 (\nu_{\rm max})^{0.75} \; ,
\label{scaling}
}
where $\nu_{\rm max}$ and $\Delta \nu$ are expressed in $\mu$Hz,
the typical width of the observed frequency range, $\delta \nu$, on both sides of $\nu_{\rm max}$ is equal to
\algn{
\delta \nu \approx  3 \Delta \nu \approx 0.84 \nu_{\rm max} ^{0.75} \; .
}
For typical observed red giant stars in case $b$, we have \smash{$50~\mu$Hz $\lesssim \nu_{\rm max} \lesssim 110~\mu$Hz} (e.g., see \figurename{}~\ref{obs}), so that we obtain
\algn{
0.26~\nu_{\rm max} \lesssim\delta \nu \lesssim  0.32~\nu_{\rm max}\; .
}
Using \eq{gamma b}, it is possible to check for the considered range of $\nu_{\rm max}$ that the remainder over the observed frequency range $\nu_{\rm max} \pm \delta \nu$ satisfies the inequality
\algn{
\left| \gamma(\nu ;\nu_{\rm max})\right| \le \left|\gamma(\nu_{\rm max}\pm\delta \nu;\nu_{\rm max})\right| \lesssim \sigma_{\varepsilon} \sim 0.08 \; .
}
As a result, the remainder is smaller than or comparable to the typical uncertainties on the observed values of the gravity offset.

Therefore, given the current precision on the measurement of these parameters, we conclude that the local asymptotic expressions of $\varepsilon_{\rm g}$ and $\Delta \Pi_1$ defined in \sectionname{}~\ref{asymptotic definition} are both sufficient and representative of the observed values. The impact of the remainder on the comparison with the observed values performed in \sectionname{}~\ref{obs} is thus negligible and does not modify the main conclusions, as mentioned before.

\subsection{Sensitivity of the results on the assumptions}
\label{discussion assumption}

The theoretical expression of $\varepsilon_{\rm g}$ relies on several simplifying hypotheses regarding the structure of the region between the hydrogen-burning shell and the base of the convective zone. Here, we propose to discuss the sensitivity of the result on these assumptions.

On the one hand, we assume that the evanescent region is thick. While this is met at first approximation for evolved red giant stars in case $b$, it is questionable for younger stars in case $a$ as shown by the observations of \cite{Mosser2017b}. \cite{Takata2016a} analytically treated the other limiting case of a very thin evanescent zone by assuming that $\mathcal{S }_1$ varies as a similar power law of radius to $\mathcal{N}$. In this hypothesis, the value of $\varepsilon_{\rm g}$ is expected to remain between 0.2 and 0.35 despite a large set of typical configurations considered for the evanescent region. This is in agreement with the observations (e.g., see gray rectangle in \figurename{}~\ref{obs}), as already mentioned by \cite{Mosser2018}. Moreover, given the small sensitivity of $\varepsilon_{\rm g}$ to the properties of the evanescent region (see Fig.~4 of T16), we expect that its variation over the observed frequency range to be much smaller than $\sigma_{\varepsilon}\sim0.08$. As a conclusion, the local asymptotic definitions of the period spacing and gravity offset remain representative of the observed values in the case of a thin evanescent region, and thus their comparison is also relevant.

On the other hand, it is obvious that the transition between cases $a$ and $b$ is more progressive than assumed in the model presented in \sectionname{}~\ref{modeling} (e.g., compare Figs.~\ref{prop_diag}~and~\ref{scheme}). In this region, it seems more reasonable to assume that the logarithmic derivative of $\mathcal{N}$, denoted with $\beta^\star$, can vary with radius, that is,
\algn{
\deriv{\ln \mathcal{N}}{r}=-\beta^{\star}(r) \; .
\label{power law 3}
}
The exponent $\beta^\star$ is supposed to be much higher than unity and to tend rapidly to infinity near $r_{\rm b}$. In this intermediate case, the estimate of the integrals in \eqs{I_20}{I_10}~can be obtained using two main assumptions. First, we can assume that the frequency dependent quantities $\beta^\star(r_2)$ and $J(r_2)$ do not vary too much over the observed frequency range around $\nu_{\rm max}$. As a result, the integral $\mathcal{I}_1$ in \eq{I_20} can be computed around $\nu_{\rm max}$ by replacing $r_{\rm b}$ with $r_2(\nu_{\rm max})$, such that the result is similar to \eq{I_2} by replacing $\mathcal{N}_{\rm b}$ with $2\pi \nu_{\rm max}$, that is,
\algn{
\mathcal{I}_1 \approx \frac{\sqrt{2}}{\sigma}\left[\int_{0}^{r_{2,{\rm max}}}  \frac{N}{r} \dd r +  \frac{2\pi \nu_{\rm max}J(r_{2,{\rm max}})}{\beta^\star(r_{2,{\rm max}})}\right] - \frac{\sqrt{2}J(r_{2,{\rm max}})}{\beta^\star(r_{2,{\rm max}})} \; ,
}
where $r_{2,{\rm max}}=r_2(\nu_{\rm max})$. Second, as $\nu_{\rm max}$ decreases and $r_2$ tends to $r_{\rm b}$, the main contribution to the integrals in \eq{I_10}~comes from the region near $r_2$ since $\mathcal{N}$ in the integrands sharply decrases with $\ln r$ at the base of the convective zone. As a result, \eq{I_10} can be crudely estimated in a similar way to \eq{I_1} by assuming that $J$ and $\beta^\star$ are constant and equal to their values at $r_2(\nu)$, that is,
\algn{
\mathcal{I}_2&\approx \frac{\sqrt{2} J(r_{2,{\rm max}})}{\beta^\star(r_{2,{\rm max}})} \left[ 1-\frac{\pi}{2}\right] \; ,
}
where the first assumptions $J\approx J(r_{2,{\rm max}})$ and $\beta^\star\approx \beta^\star(r_{2,{\rm max}})$ over the observed frequency range were used. Following the same reasoning as in \sectionname{}~\ref{case $a$}, the local asymptotic expression of the gravity offset is thus similar to \eq{epsilon_g a}, that is,%
\algn{
\varepsilon_{\rm g}\approx\frac{\sqrt{2} J(r_{2,{\rm max}})}{2\beta^\star(r_{2,{\rm max}})} - \frac{1}{4} \; ,
\label{epsilon_g inter}
}
except that the relation $\beta=3J/2$ in the previous picture does not hold true in this case between $\beta^\star$ and $J$. Nevertheless, \smash{$J(r_{2,{\rm max}}) \sim 1$} in evolved red giant stars close to the transition, so that the evolution of $\varepsilon_{\rm g}$ depends only on the evolution of $\beta^\star(r_{2,{\rm max}})$. This latter parameter is equivalent to the frequency dependent arcsine function in \eq{epsilon_g b}. Because of the sharp kink of the Brunt-Väisälä frequency, the increase in $\beta^\star(r_{2,{\rm max}})$ due to the decrease in $\nu_{\rm max}$ can be assumed to be so rapid close to the transition that it can mimic the sharp decrease in the value of $\varepsilon_{\rm g}$ that was predicted by the model developed in \sectionname{}~\ref{evolution on RGB}. Using \eq{epsilon_g inter}, the rapid decrease in $\varepsilon_{\rm g}$ resulting from the change in $\beta^\star$ from $3/2$ to infinity is equal to \smash{$\sqrt{2}/3 \approx 0.47$}, which reasonably agrees with the observations in \figurename{}~\ref{obs} given the crude assumptions used to derive \eq{epsilon_g inter}. In these considerations, the same conclusions are therefore reached, the drop in the value of $\varepsilon_{\rm g}$ results from the kink in the Brunt-Väisälä frequency near the turning point $r_2$ at the transition.

In summary, despite the simple modeling we used, it appears to be sufficient for our purpose and does not impact our interpretation of the observed global variations in $\varepsilon_{\rm g}$ in terms of internal structure.

\subsection{Simultaneous variations in $\Delta \Pi_1$ and $q$}
\label{DP q}

While $\varepsilon_{\rm g}$ exhibits a rapid characteristic decrease for stars at the transition between cases $a$ and $b$, we can wonder whether simultaneous rapid variations in the period-spacing $\Delta \Pi_1$ and in the coupling factor $q$ of mixed modes are observable or not.

Regarding $\Delta \Pi_1$, \eqs{Delta Pi YR}{Delta Pi ER}~also predict a sharp increase at the transition between cases $a$ and $b$. Indeed, $\Delta \Pi_1$ is expected to decrease smoothly for stars in case $a$ owing to the increase both in the integral of the Brunt-Väisälä frequency between 0 and $r_{\rm b}$ and in the parameter $\mathcal{N}_{\rm b}$ during evolution. After the transition, an extra term appears in \eq{Delta Pi ER} depending on $\nu_{\rm max}$. As a result, the derivative of the period spacing with respect to $\nu_{\rm max}$ is discontinuous and tends to infinity at the transition, whence an expected sharp increase during the evolution. The maximum relative amplitude of the jump in $\Delta \Pi_1$ predicted by the model can be estimated considering that the ratio $2\pi \nu_{\rm max}/\mathcal{N}_{\rm b}$ varies quasi-instantaneously from unity to zero at the transition while $N$ and $\mathcal{N}_{\rm b}$ remain about unmodified. According to \eqs{Delta Pi YR}{Delta Pi ER}, and the fact that $\mathcal{N}_{\rm b} \sim 2 \pi \nu_{\rm max} $ at the transition, the relative rapid variation in the period spacing at the transition is thus expected to be lower than
\algn{
\frac{\delta( \Delta \Pi_1)_{a \rightarrow b}}{\Delta \Pi_1} \lesssim \frac{2 \mathcal{N}_{\rm b} }{3}\frac{\sqrt{2}}{2 \pi^2} \Delta \Pi_1 \sim \frac{2 \sqrt{2}}{3 \pi} \nu_{\rm max} \Delta \Pi_1 \; .
\label{jump delta}
}
This transition occurs for stars around $\nu_{\rm max} \approx 80~\mu$Hz, which corresponds to $\Delta \Pi_1 \approx 75 $ s \citep[e.g.,][]{Vrard2016}.
Following \eq{jump delta}, this is equivalent to a rapid relative increase smaller than 0.15\% in $\Delta \Pi_1$. This variation is too low to be clearly detected since it is on the
order of the relative precision on the period spacing for these stars \citep{Mosser2018}. This implies that $\varepsilon_{\rm g}$ is more sensitive than $\Delta \Pi_1$ to the transition, which is reasonable since $\varepsilon_{\rm g}$ corresponds to the (small) higher order term in the phase term $\Theta_{\rm g}$ while $\Delta \Pi_1$ represents the leading-order term that is mainly sensitive to the properties of the region in the middle of the buoyancy cavity.

Besides, the coupling factor depends on the wave transmission factor through the evanescent region and thus on the properties of this region \citep[e.g.,][]{Takata2016b}. The evanescent region is located between the profiles of $\mathcal{N}$ and $\mathcal{S}_1$, that is, between turning points $r_2$ and $r_3$ for which $\sigma = \mathcal{N}(r_2)$ and $\sigma=\mathcal{S}_1(r_3)$. Without going into the details, we thus understand that the coupling factor is also likely to be sensitive to the kink in the Brunt-Väisälä frequency at the transition between cases $a$ and $b$ since it significantly modifies the structure of the evanescent region (e.g., see \figurename{}~\ref{prop_diag}). The factor $q$ could therefore potentially present a noticeable change of trend during evolution that would be associated with this transition. For the considered stars, that is, for $50 ~\mu$Hz~$\lesssim \nu_{\rm max}\lesssim 110~\mu$Hz, we did not perceive any obvious change of behavior as a function of $\nu_{\rm max}$ in the observed values of $q$ obtained by \cite{Mosser2017b}. The lack of data for $\nu_{\rm max} \lesssim 80~\mu$Hz and the current uncertainties on the measurements partly make the search for such a feature difficult. Analytical studies of the coupling factor around the transition will be useful to guide future observational investigations but need for further theoretical efforts. Indeed, in addition to consider the effect of the kink in the Brunt-Väisälä frequency on the variation of $q$ with evolution, the analysis will also have to deal with the influence of the spike in the Brunt-Väisälä frequency and of the discontinuity in the Lamb frequency at the base of the convective region resulting from the presence of a sharp molecular weight gradient (see \sectionname{}~\ref{mu gradient}). This will demand going beyond the picture given by the current asymptotic analyses of mixed modes provided by \cite{Shibahashi1979} and \cite{Takata2016a}.

Therefore, we conclude that neither $\Delta \Pi_1$ nor $q$ can currently bring us a signature on the transition between cases $a$ and $b$ as clear as that observed in $\varepsilon_{\rm g}$. More detailed analyses should be performed in the future to find evidence of a given signature associated with the transition in these parameters.

\subsection{Possible effect of the $\mu$-gradient at the base of the convective zone}
\label{mu gradient}

Figure~\ref{prop_diag} shows that a spike in the (modified) Brunt-Väisälä frequency exists at the base of the convective zone for evolved red giant stars in case $b$. This is the consequence of a strong molecular weight gradient (or $\mu$-gradient). This $\mu$-gradient causes the luminosity bump in the later stage when the hydrogen-burning shell reaches the associated zone. Since the luminosity bump is observationally detected \citep[e.g.,][]{Khan2018}, there is no doubt on the existence of this $\mu$-gradient in real stars.

Such a sharp variation can slightly perturb the oscillation period and create a so-called buoyancy glitch \citep[e.g.,][]{Brassard1992,Miglio2008,Cunha2015}. 
Indeed, as shown by \cite{Cunha2015}, the presence of a sharp $\mu$-gradient inside the buoyancy cavity locally results in a discontinuity in the wave function derivative with respect to radius, leading to an additional phase term in the expression of $\Theta_{\rm g}$ in \eq{Theta_g final}. It is thus obvious that this feature can modify the physical interpretation of the observed values of the gravity offset in case $b$ if its contribution to $\Theta_{\rm g}$ is of the same order as the asymptotic predictions obtained without considering the glitch. We propose to discuss its impact in a simple way focusing on stars that have just spent the transition, that is, for which $2\pi \nu_{\rm max} \sim \mathcal{N}_{\rm b}$. Indeed, most of the observed values of the gravity offset in case $b$ were obtained for $50~\mu$Hz~$\lesssim \nu_{\rm max} \lesssim 110~\mu$Hz (e.g., see \figurename{}~\ref{obs}), which corresponds to the $\nu_{\rm max}$ frequency range of the stars just after the transition that is expected from stellar models (e.g., see \figurename{}~\ref{prop_diag}).
To go further, we adapt the formulation of buoyancy glitches given by \citet[][hereafter, C15]{Cunha2015}. In other words, we assume that the spike in $N^2$ can be modeled by a Dirac $\delta$ function located at $r_\star$ and with an amplitude $A_\star$. However, in contrast to C15, we do not assume that this sharp feature is located far away from the turning point $r_2$. Instead, we assume that $r_\star$ is very close to $r_2$, where the radial wavenumber vanishes. Under these considerations, the discontinuity in the wave function derivative with respect to radius at $r_\star$ that is provided by Eq.~(11) of C15 becomes
\algn{
\left[\deriv{\Phi_{\rm in}}{r}-\deriv{\Phi_{\rm out}}{r} \right]_{r_\star} = A_\star k_r^2(r_\star) \Phi_{\rm in}(r_\star) \approx 0 \; ,
\label{continuity}
}
where $\Phi_{\rm in}(r)$ and $\Phi_{\rm out}(r)$ are the wave functions below and above $r_\star$, respectively. Equation~(\ref{continuity}) shows that the derivative of the wave function remains continuous at first approximation near $r_\star$. This holds true whatever the amplitude of the $\mu$-gradient since $k_r(r_\star) \approx k_r (r_2) \equiv 0$ in the considered hypothesis. As a result, the contribution of the additional phase term associated with the glitch to $\Theta_{\rm g}$ must be of higher order than $\varepsilon_{\rm g}$.

In conclusion, we find that the impact of the spike in the Brunt-Väisälä frequency at the base of the convective zone is small at first approximation and does not modify the interpretation of the global observed variations of the gravity offset in evolved red giant stars. This result will have to be confirmed by more detailed analyses based either on theoretical developments or stellar models, accounting both for a more realistic shape of the spike in $N$ and for slightly different values between $r_\star$ and $r_2$, as observed in \figurename{}~\ref{prop_diag}. This must be subject to future works, which are beyond the scope of this paper.

\subsection{Overshooting below the convective region}
\label{overshooting}

To match the observed location of the luminosity bump in low-mass stars with stellar models, \cite{Khan2018} included overshooting below the convective envelope. They demonstrated that an adiabatic extent of the convective zone of about $0.3H$ on average over the mass and metallicity ranges of their sample, where $H$ is the pressure scale height, is sufficient to fit the observations. This work gives a clear evidence that overshooting at the base of the convective region must be taken into account in red giant stars and thus has the potential to modify the value of the gravity offset.

However, while overshooting decreases the radius of the base of the convective region $r_{\rm b}$, it does not change the formal profile of the (modified) Brunt-Väisälä frequency just below $r_{\rm b}$. Indeed, it is possible to check in typical red giant models including overshooting that $\mathcal{N}$ still varies at first approximation as a power law of radius between the hydrogen-burning shell and the convective envelope before sharply vanishing close to $r_{\rm b}$. Therefore, the assumptions used in \sectionname{}~\ref{modeling} remain valid and the analytical expressions of the gravity offset provided in \eqs{epsilon_g a}{epsilon_g b}~are not modified in the presence of overshooting.

For young red giant stars in case $a$, the expression of $\varepsilon_{\rm g}$ in \eq{epsilon_g a} is independent of the internal structure, so that overshooting has no effect on the value of the gravity offset predicted by the model. For more evolved stars in case $b$, the expression of $\varepsilon_{\rm g}$ in \eq{epsilon_g a} depends on the internal structure only through the ratio $2\pi \nu_{\rm max} / \mathcal{N}_{\rm b}$. Therefore, the effect of overshooting on $\varepsilon_{\rm g}$ at a given value of $\nu_{\rm max}$ can only result from its impact on $\mathcal{N}_{\rm b}$. Since the base of the convective region is deeper with overshooting, the value of $\mathcal{N}_{\rm b}$ is higher according to \eq{Nb}. As shown in \appendixname{}~\ref{Nb evol}, this trend is well reproduced in stellar models. Moreover, it turns out that the variation rate of $\mathcal{N}_{\rm b}$ as a function of $\nu_{\rm max}$ before the luminosity bump is comparable with or without overshooting. As a result, at a given value of $\nu_{\rm max}$, overshooting leads only to higher $\nu_{\rm max,t}$ at the transition between cases $a$ and $b$ (i.e., for which $2\pi \nu_{\rm max, t} \approx \mathcal{N}_{\rm b}$). For instance, we found that the inclusion of an adiabatic overshoot region of about $0.3 H$ in a $1.2M_\odot$ evolutionary sequence with the same input physics as described in \sectionname{}~\ref{unclear} leads to an increase in $\nu_{\rm max,t}$ from about $95~\mu$Hz to about $115~\mu$Hz.

As a conclusion, the effect of overshooting is already included in the analytical model through the parameter $\nu_{\rm max,t}$. Indeed, as shown by simple considerations and stellar models, the thicker the overshooting region, the higher the value of $\nu_{\rm max,t}$ since the lower the value of $r_{\rm b}$ (see \appendixname{}~\ref{Nb evol}). Moreover, for reasonable values of the overshoot thickness based on the study of the luminosity bump \citep{Khan2018}, it appears that its effect tends to reinforce the agreement between the model and observations, as mentioned in \sectionname{}~\ref{obs}

\subsection{Seismic diagnoses with $\varepsilon_{\rm g}$}

In \sectionname{}~\ref{evolution on RGB}, we demonstrated that the global variation in the gravity offset during the evolution exhibits a clear signature on the RGB that is related to the kink in the Brunt-Väisälä frequency at the interface between the radiative and convective zones. Moreover, our analytical model also provides a first insight into the potential of the gravity offset of mixed modes to probe this region in more detail. 

Indeed, on the one hand, for young red giant stars in case~$a$, the model predicts a constant value of $0.22$ for $\varepsilon_{\rm g}$ in the weak coupling limit. Nevertheless, in the strong coupling limit, \cite{Takata2016a} showed that the value of $\varepsilon_{\rm g}$ depends both on the slope of $\mathcal{N}$ in this region and on the thickness of the evanescent region. The structure of this region is ruled by the high density contrast between the core and envelope, which depends on the core contraction. The gravity offset in case $a$ has therefore the potential to inform us of the contraction rate of the helium core.
On the other hand, for more evolved red giant stars in case $b$, the model predicts that the value of $\varepsilon_{\rm g}$ depends only on $\mathcal{N}_{\rm b}$ at a given value of $\nu_{\rm max}$. As discussed in \sectionname{}~\ref{comparison} and in \appendixname{}~\ref{Nb evol}, the value of $\mathcal{N}_{\rm b}$ mainly depends on the location of the base of the convective region. Therefore, its value may inform us of the migration of the convective boundary with time, and in particular around the convective first dredge-up, and on overshooting in this region.

In practice, despite the good precision on the measurement of the gravity offset, which results from a precision on the measurement of $\Delta \Pi_1$ better than 0.5\%, the uncertainties on its value can appear at first sight to be too large to provide us with detailed constraints on the region close to $r_2$ for a given star in case $a$. Indeed, as mentioned in \sectionname{}~\ref{modeling}, the sensitivity of $\varepsilon_{\rm g}$ to the structural properties is on the
order of the error bars for stars in case $a$ according to the analytical result of \cite{Takata2016a} (e.g., see the dark gray rectangle in \figurename{}~\ref{obs}). In contrast, for more evolved stars after the transition, the observed variations in \smash{$\varepsilon_ {\rm g}^{\rm obs}$} can certainly give us more detailed insights into the internal properties. To illustrate this point, we can consider for instance a $1.2M_\odot$ star with $\nu_{\rm max} \approx 60~ \mu$Hz and $\varepsilon_{\rm g} =-0.1$. To explain this value, \eqref{epsilon_g b} requires $2\pi \nu_{\rm max} /\mathcal{N}_{\rm b} \approx 0.5$. As previously discussed in \sectionname{}~\ref{comparison} and \sectionname{}~\ref{overshooting}, such a value of $\mathcal{N}_{\rm b}$ at $\nu_{\rm max} \approx 60 ~\mu$Hz can result either from the progressive increase of $\mathcal{N}_{\rm b}$ owing to the inward migration of the convective boundary or from a slightly higher value of $\nu_{\rm max,t}$ caused, for instance, by overshooting. Indeed, without overshooting, we estimated through stellar models that $\nu_{\rm max ,t}\approx 95~\mu$Hz for a stellar mass of $1.2M_\odot$. In order to fit the observed value at $\nu_{\rm max}=60~\mu$Hz with the analytical expression in \eq{epsilon_g b}, given the assumption that $\mathcal{N}_{\rm b} \propto\nu_{\rm max}^{-x}$, we find that a value of $x \approx-0.5$ is needed. In contrast, if $\mathcal{N}_{\rm b}$ is considered as a constant, we find that an increase in the value of $\nu_{\rm max,t}$ to about $120~\mu$Hz is needed for the prediction of the analytical model to match the given value of $\varepsilon_{\rm g}$ at $\nu_{\rm max}=60~\mu$Hz. Stellar models show that this can be achieved with an adiabatic extent of the convective region of about $0.3H$. In reality, such a value of $\varepsilon_{\rm g}$ at $\nu_{\rm max}=60~\mu$Hz may result from both effects. Equivalently, we note that the different picture considered for stars close to the transition in \sectionname{}~\ref{modeling} requires $\beta^\star(r_{2,{\rm max}})\sim 5$ in \eqref{epsilon_g inter} (with $J(r_{2,{\rm max}}) \approx 1$) to explain $\varepsilon_{\rm g}=-0.1$. In this view, the slope of the Brunt-Väisälä at $r_{2,{\rm max}}$ plays the same role as the ratio $2\pi \nu_{\rm max}/\mathcal{N}_{\rm b}$ in \eq{epsilon_g b} and thus may be sensitive at a given value of $\nu_{\rm max}$ to overshooting and the migration speed of the base of the convective zone.

The analytical model that was developed therefore supports the gravity offset as a promising source of information on the base of the convective region. Further studies will have to be performed to assess the ability of this parameter to constrain stellar models. Refinement in the analytical model that was developed in this work, fully accounting for instance for a realistic shape of the Brunt-Väisälä frequency, the sharp $\mu$-gradient at the base of the convective region, or a better representation of the evolution of $\mathcal{N}_{\rm b}$ as a function of $\nu_{\rm max}$ , will help guide such investigations.

%
\section{Conclusions}
\label{conclusions}

In this paper, we addressed from a theoretical point of view the variations in the gravity offset of mixed modes that were observed on the RGB by \cite{Mosser2018}. 
After having properly introduced the notion of gravity offset, we considered the asymptotic analyses of mixed modes by \cite{Shibahashi1979}, \cite{Takata2016a} and \cite{Takata2016b} to make the link between the internal properties of stars and the observed values. First, we clearly showed in general terms that this parameter is related to the properties of the boundaries of the buoyancy cavity through the contributions of the phase lags introduced at reflection and of the wavenumber integral in the vicinity of the turning points (i.e., close to $r_1$ and $r_2$). We also demonstrated that its value must remain on the
order of unity. Then, by modeling at first approximation the Brunt-Väisälä frequency as a power law of radius between the hydrogen-burning shell and the base of the convective zone, we obtained simplified local asymptotic expressions for the period spacing and gravity offset, denoted with $\Delta \Pi_1$ and $\varepsilon_{\rm g}$, respectively, over a narrow frequency range around $\nu_{\rm max}$. These local asymptotic values were shown to be representative of the observed values. As a result, the observed quantities could be analyzed within the asymptotic framework.

The considered model predicts a constant value for $\varepsilon_{\rm g}$ close to 0.22 in the first part of the ascent of the RGB. This peculiar value is the result of the high density contrast between the helium core and the base of the convective envelope. At a peculiar instant just before the luminosity bump, it is expected to drop rapidly by a mean value of about $0.2$ in the directly following stages. This fall in the value of the gravity offset is a consequence of the kink of the Brunt-Väisälä frequency near the base of the convective region when the outer turning point associated with the buoyancy cavity becomes and remains very close to this latter region during the subsequent evolution on the RGB evolution. Using stellar models, we estimated that this transition must occur around $50~\mu$Hz$\lesssim\nu_{\rm max} \lesssim 110~\mu$Hz for a typical mass range of the observed red giant stars. These predictions are in global agreement with the observed variations. As a result, we conclude that the drop observed in the value of the gravity offset in red giant stars is the signature of the kink of the Brunt-Väisälä frequency at the transition between the radiative and convective regions. The simplified assumptions on the structure of the upper layers of the radiative zone that were considered in the model were checked a posteriori not to modify the conclusions. For more evolved stars, for which the visibility of gravity-dominated mixed modes is too low to be observed, the model predicts a lower limit of -0.25. 

The agreement between theoretical predictions and global trend in the observations of the gravity offset validates both the use of the asymptotic expressions to interpret the data and the relevancy of the measurements. We demonstrated that no clear signature of the transition such as that observed in $\varepsilon_{\rm g}$ is currently expected to be detected in the period spacing and the coupling factor of mixed modes. However, both seismic parameters certainly remains sensitive to this transition. More thorough analyses of the data guided by new theoretical developments will be needed in the future to detect a signature associated with the transition in these parameters. This will provide additional support for the relevancy of the present conclusions regarding $\varepsilon_{\rm g}$. 

To conclude, this work represents a first step toward a methodic exploitation of the gravity offset for the characterization of red giant stars, and in particular for the study of the region located between the helium core and the base of the convective envelope. The ability of this parameter to probe the profile of the Brunt-Väisälä frequency in this region allows us to constrain the contraction rate of the helium core, the migration of the base of the convective zone, and the efficiency and the extent of the convective boundary mixing in this region. In this work, we showed in particular that accounting for these ingredients tends to improve the agreement between the predictions of the analytical model and the observations. This promising potential calls for further investigations. The next step will directly consist in extending this study to red clump stars. The analysis will have to carefully take into account the presence of a convective core in these central helium-burning stars, which will be subject of a future work.

\begin{acknowledgements}
C. P. especially thank M.-J. Goupil and K. Belkacem for their fruitful comments on the paper and their unwavering support. The authors also thank the entire {\it Kepler} team, whose efforts
made these results possible, as well as the anonymous referee whose comments greatly helped improve the final version of the paper. M.~T. is grateful to Paris Observatory for the support of his stay at the observatory. 
During this work, C. P. was partially funded by a postdoctoral research grants from CNRS (France) and from F.R.S.-FNRS (Belgium). M. T. is partially supported by JSPS KAKENHI Grant Number 18K03695.
We appreciate the financial support from the Programme National de Physique Stellaire
(CNRS/INSU). B. M. acknowledges the support of the International Space Institute (ISSI) for the program AsteroSTEP (Asteroseismology of STEllar Populations).
\end{acknowledgements}

\bibliographystyle{aa} 
\bibliography{bib} 

\begin{thebibliography}{45}
\expandafter\ifx\csname natexlab\endcsname\relax\def\natexlab#1{#1}\fi

\bibitem[{{Abramowitz} \& {Stegun}(1972)}]{Abramowitz1972}
{Abramowitz}, M. \& {Stegun}, I.~A. 1972, {Handbook of Mathematical Functions}

\bibitem[{{Asplund} {et~al.}(2009){Asplund}, {Grevesse}, {Sauval}, \&
  {Scott}}]{AGS2009}
{Asplund}, M., {Grevesse}, N., {Sauval}, A.~J., \& {Scott}, P. 2009, \araa, 47,
  481

\bibitem[{{Baglin} {et~al.}(2006{\natexlab{a}}){Baglin}, {Auvergne}, {Barge},
  {Deleuil}, {Catala}, {Michel}, {Weiss}, \& {COROT Team}}]{Baglin2006a}
{Baglin}, A., {Auvergne}, M., {Barge}, P., {et~al.} 2006{\natexlab{a}}, in ESA
  Special Publication, Vol. 1306, ESA Special Publication, ed. M.~{Fridlund},
  A.~{Baglin}, J.~{Lochard}, \& L.~{Conroy}, 33

\bibitem[{{Baglin} {et~al.}(2006{\natexlab{b}}){Baglin}, {Auvergne},
  {Boisnard}, {Lam-Trong}, {Barge}, {Catala}, {Deleuil}, {Michel}, \&
  {Weiss}}]{Baglin2006b}
{Baglin}, A., {Auvergne}, M., {Boisnard}, L., {et~al.} 2006{\natexlab{b}}, in
  COSPAR Meeting, Vol.~36, 36th COSPAR Scientific Assembly, 3749

\bibitem[{{Bedding} {et~al.}(2011){Bedding}, {Mosser}, {Huber},
  {Montalb{\'a}n}, {Beck}, {Christensen-Dalsgaard}, {Elsworth},
  {Garc{\'{\i}}a}, {Miglio}, {Stello}, {White}, {De Ridder}, {Hekker}, {Aerts},
  {Barban}, {Belkacem}, {Broomhall}, {Brown}, {Buzasi}, {Carrier}, {Chaplin},
  {di Mauro}, {Dupret}, {Frandsen}, {Gilliland}, {Goupil}, {Jenkins},
  {Kallinger}, {Kawaler}, {Kjeldsen}, {Mathur}, {Noels}, {Silva Aguirre}, \&
  {Ventura}}]{Bedding2011}
{Bedding}, T.~R., {Mosser}, B., {Huber}, D., {et~al.} 2011, \nat, 471, 608

\bibitem[{{Belkacem} {et~al.}(2013){Belkacem}, {Samadi}, {Mosser}, {Goupil}, \&
  {Ludwig}}]{Belkacem2013}
{Belkacem}, K., {Samadi}, R., {Mosser}, B., {Goupil}, M.-J., \& {Ludwig}, H.-G.
  2013, in Astronomical Society of the Pacific Conference Series, Vol. 479,
  Progress in Physics of the Sun and Stars: A New Era in Helio- and
  Asteroseismology, ed. H.~{Shibahashi} \& A.~E. {Lynas-Gray}, 61

\bibitem[{{Borucki} {et~al.}(2010){Borucki}, {Koch}, {Basri}, {Batalha},
  {Brown}, {Caldwell}, {Caldwell}, {Christensen-Dalsgaard}, {Cochran},
  {DeVore}, {Dunham}, {Dupree}, {Gautier}, {Geary}, {Gilliland}, {Gould},
  {Howell}, {Jenkins}, {Kondo}, {Latham}, {Marcy}, {Meibom}, {Kjeldsen},
  {Lissauer}, {Monet}, {Morrison}, {Sasselov}, {Tarter}, {Boss}, {Brownlee},
  {Owen}, {Buzasi}, {Charbonneau}, {Doyle}, {Fortney}, {Ford}, {Holman},
  {Seager}, {Steffen}, {Welsh}, {Rowe}, {Anderson}, {Buchhave}, {Ciardi},
  {Walkowicz}, {Sherry}, {Horch}, {Isaacson}, {Everett}, {Fischer}, {Torres},
  {Johnson}, {Endl}, {MacQueen}, {Bryson}, {Dotson}, {Haas}, {Kolodziejczak},
  {Van Cleve}, {Chandrasekaran}, {Twicken}, {Quintana}, {Clarke}, {Allen},
  {Li}, {Wu}, {Tenenbaum}, {Verner}, {Bruhweiler}, {Barnes}, \&
  {Prsa}}]{Borucki2010}
{Borucki}, W.~J., {Koch}, D., {Basri}, G., {et~al.} 2010, Science, 327, 977

\bibitem[{{Brassard} {et~al.}(1992){Brassard}, {Fontaine}, {Wesemael}, \&
  {Hansen}}]{Brassard1992}
{Brassard}, P., {Fontaine}, G., {Wesemael}, F., \& {Hansen}, C.~J. 1992, \apjs,
  80, 369

\bibitem[{{Brekhovskikh}(1980)}]{Brekhovskikh1980}
{Brekhovskikh}, L.~M. 1980, in Applied Mathematics and Mechanics, Vol.~16,
  Waves in Layered Media, ed. L.~Brekhovskikh (Elsevier), 161 -- 224

\bibitem[{{Buysschaert} {et~al.}(2016){Buysschaert}, {Beck}, {Corsaro},
  {Christensen-Dalsgaard}, {Aerts}, {Arentoft}, {Kjeldsen}, {Garc{\'{\i}}a},
  {Silva Aguirre}, \& {Degroote}}]{Buysschaert2016}
{Buysschaert}, B., {Beck}, P.~G., {Corsaro}, E., {et~al.} 2016, \aap, 588, A82

\bibitem[{{Chaplin} \& {Miglio}(2013)}]{Chaplin2013}
{Chaplin}, W.~J. \& {Miglio}, A. 2013, \araa, 51, 353

\bibitem[{{Christensen-Dalsgaard}(2015)}]{JCD2015}
{Christensen-Dalsgaard}, J. 2015, \mnras, 453, 666

\bibitem[{{Cowling}(1941)}]{Cowling1941}
{Cowling}, T.~G. 1941, \mnras, 101, 367

\bibitem[{{Cunha} {et~al.}(2015){Cunha}, {Stello}, {Avelino},
  {Christensen-Dalsgaard}, \& {Townsend}}]{Cunha2015}
{Cunha}, M.~S., {Stello}, D., {Avelino}, P.~P., {Christensen-Dalsgaard}, J., \&
  {Townsend}, R.~H.~D. 2015, \apj, 805, 127

\bibitem[{{de Assis Peralta} {et~al.}(2018){de Assis Peralta}, {Samadi}, \&
  {Michel}}]{Peralta2018}
{de Assis Peralta}, R., {Samadi}, R., \& {Michel}, E. 2018, Astronomische
  Nachrichten, 339, 134

\bibitem[{{Deheuvels} {et~al.}(2014){Deheuvels}, {Do{\u g}an}, {Goupil},
  {Appourchaux}, {Benomar}, {Bruntt}, {Campante}, {Casagrande}, {Ceillier},
  {Davies}, {De Cat}, {Fu}, {Garc{\'{\i}}a}, {Lobel}, {Mosser}, {Reese},
  {Regulo}, {Schou}, {Stahn}, {Thygesen}, {Yang}, {Chaplin},
  {Christensen-Dalsgaard}, {Eggenberger}, {Gizon}, {Mathis},
  {Molenda-{\.Z}akowicz}, \& {Pinsonneault}}]{Deheuvels2014}
{Deheuvels}, S., {Do{\u g}an}, G., {Goupil}, M.~J., {et~al.} 2014, \aap, 564,
  A27

\bibitem[{{Di Mauro}(2016)}]{DiMauro2016}
{Di Mauro}, M.~P. 2016, in Frontier Research in Astrophysics II, held 23-28
  May, 2016 in Mondello (Palermo), Italy (FRAPWS2016). Online at
  <https://pos.sissa.it/cgi-bin/reader/conf.cgi?confid=269>, id.29, 29

\bibitem[{{Dupret} {et~al.}(2009){Dupret}, {Belkacem}, {Samadi}, {Montalban},
  {Moreira}, {Miglio}, {Godart}, {Ventura}, {Ludwig}, {Grigahc{\`e}ne},
  {Goupil}, {Noels}, \& {Caffau}}]{Dupret2009}
{Dupret}, M.-A., {Belkacem}, K., {Samadi}, R., {et~al.} 2009, \aap, 506, 57

\bibitem[{{Gehan} {et~al.}(2018){Gehan}, {Mosser}, {Michel}, {Samadi}, \&
  {Kallinger}}]{Gehan2018}
{Gehan}, C., {Mosser}, B., {Michel}, E., {Samadi}, R., \& {Kallinger}, T. 2018,
  \aap, 616, A24

\bibitem[{{Grosjean} {et~al.}(2014){Grosjean}, {Dupret}, {Belkacem},
  {Montalban}, {Samadi}, \& {Mosser}}]{Grosjean2014}
{Grosjean}, M., {Dupret}, M.-A., {Belkacem}, K., {et~al.} 2014, \aap, 572, A11

\bibitem[{{Hekker} \& {Christensen-Dalsgaard}(2017)}]{Hekker2017}
{Hekker}, S. \& {Christensen-Dalsgaard}, J. 2017, \aapr, 25, 1

\bibitem[{{Hekker} {et~al.}(2018){Hekker}, {Elsworth}, \&
  {Angelou}}]{Hekker2018}
{Hekker}, S., {Elsworth}, Y., \& {Angelou}, G.~C. 2018, \aap, 610, A80

\bibitem[{{Khan} {et~al.}(2018){Khan}, {Hall}, {Miglio}, {Davies}, {Mosser},
  {Girardi}, \& {Montalb{\'a}n}}]{Khan2018}
{Khan}, S., {Hall}, O.~J., {Miglio}, A., {et~al.} 2018, \apj, 859, 156

\bibitem[{{Kippenhahn} {et~al.}(2012){Kippenhahn}, {Weigert}, \&
  {Weiss}}]{Kippenhahn2012}
{Kippenhahn}, R., {Weigert}, A., \& {Weiss}, A. 2012, {Stellar Structure and
  Evolution}

\bibitem[{{Kjeldsen} \& {Bedding}(1995)}]{Kjeldsen1995}
{Kjeldsen}, H. \& {Bedding}, T.~R. 1995, \aap, 293, 87

\bibitem[{{Lagarde} {et~al.}(2016){Lagarde}, {Bossini}, {Miglio}, {Vrard}, \&
  {Mosser}}]{Lagarde2016}
{Lagarde}, N., {Bossini}, D., {Miglio}, A., {Vrard}, M., \& {Mosser}, B. 2016,
  \mnras, 457, L59

\bibitem[{{Marques} {et~al.}(2013){Marques}, {Goupil}, {Lebreton}, {Talon},
  {Palacios}, {Belkacem}, {Ouazzani}, {Mosser}, {Moya}, {Morel}, {Pichon},
  {Mathis}, {Zahn}, {Turck-Chi{\`e}ze}, \& {Nghiem}}]{Marques2013}
{Marques}, J.~P., {Goupil}, M.~J., {Lebreton}, Y., {et~al.} 2013, \aap, 549,
  A74

\bibitem[{{Miglio} {et~al.}(2008){Miglio}, {Montalb{\'a}n}, {Noels}, \&
  {Eggenberger}}]{Miglio2008}
{Miglio}, A., {Montalb{\'a}n}, J., {Noels}, A., \& {Eggenberger}, P. 2008,
  \mnras, 386, 1487

\bibitem[{{Montalb{\'a}n} {et~al.}(2013){Montalb{\'a}n}, {Miglio}, {Noels},
  {Dupret}, {Scuflaire}, \& {Ventura}}]{Montalban2013}
{Montalb{\'a}n}, J., {Miglio}, A., {Noels}, A., {et~al.} 2013, \apj, 766, 118

\bibitem[{{Mosser} {et~al.}(2014){Mosser}, {Benomar}, {Belkacem}, {Goupil},
  {Lagarde}, {Michel}, {Lebreton}, {Stello}, {Vrard}, {Barban}, {Bedding},
  {Deheuvels}, {Chaplin}, {De Ridder}, {Elsworth}, {Montalban}, {Noels},
  {Ouazzani}, {Samadi}, {White}, \& {Kjeldsen}}]{Mosser2014}
{Mosser}, B., {Benomar}, O., {Belkacem}, K., {et~al.} 2014, \aap, 572, L5

\bibitem[{{Mosser} {et~al.}(2012{\natexlab{a}}){Mosser}, {Elsworth}, {Hekker},
  {Huber}, {Kallinger}, {Mathur}, {Belkacem}, {Goupil}, {Samadi}, {Barban},
  {Bedding}, {Chaplin}, {Garc{\'{\i}}a}, {Stello}, {De Ridder}, {Middour},
  {Morris}, \& {Quintana}}]{Mosser2012b}
{Mosser}, B., {Elsworth}, Y., {Hekker}, S., {et~al.} 2012{\natexlab{a}}, \aap,
  537, A30

\bibitem[{{Mosser} {et~al.}(2018){Mosser}, {Gehan}, {Belkacem}, {Samadi},
  {Michel}, \& {Goupil}}]{Mosser2018}
{Mosser}, B., {Gehan}, C., {Belkacem}, K., {et~al.} 2018, \aap, 618, A109

\bibitem[{{Mosser} {et~al.}(2012{\natexlab{b}}){Mosser}, {Goupil}, {Belkacem},
  {Michel}, {Stello}, {Marques}, {Elsworth}, {Barban}, {Beck}, {Bedding}, {De
  Ridder}, {Garc{\'{\i}}a}, {Hekker}, {Kallinger}, {Samadi}, {Stumpe},
  {Barclay}, \& {Burke}}]{Mosser2012a}
{Mosser}, B., {Goupil}, M.~J., {Belkacem}, K., {et~al.} 2012{\natexlab{b}},
  \aap, 540, A143

\bibitem[{{Mosser} {et~al.}(2013){Mosser}, {Michel}, {Belkacem}, {Goupil},
  {Baglin}, {Barban}, {Provost}, {Samadi}, {Auvergne}, \&
  {Catala}}]{Mosser2013b}
{Mosser}, B., {Michel}, E., {Belkacem}, K., {et~al.} 2013, \aap, 550, A126

\bibitem[{{Mosser} \& {Miglio}(2016)}]{Mosser2016}
{Mosser}, B. \& {Miglio}, A. 2016, in The CoRoT Legacy Book: The Adventure of
  the Ultra High Precision Photometry from Space, ed. {CoRot Team}, 197

\bibitem[{{Mosser} {et~al.}(2017){Mosser}, {Pin{\c c}on}, {Belkacem}, {Takata},
  \& {Vrard}}]{Mosser2017b}
{Mosser}, B., {Pin{\c c}on}, C., {Belkacem}, K., {Takata}, M., \& {Vrard}, M.
  2017, \aap, 600, A1

\bibitem[{{Pin{\c c}on} {et~al.}(2017){Pin{\c c}on}, {Belkacem}, {Goupil}, \&
  {Marques}}]{Pincon2017}
{Pin{\c c}on}, C., {Belkacem}, K., {Goupil}, M.~J., \& {Marques}, J.~P. 2017,
  \aap, 605, A31

\bibitem[{{Provost} \& {Berthomieu}(1986)}]{Provost1986}
{Provost}, J. \& {Berthomieu}, G. 1986, \aap, 165, 218

\bibitem[{{Shibahashi}(1979)}]{Shibahashi1979}
{Shibahashi}, H. 1979, \pasj, 31, 87

\bibitem[{{Takata}(2006)}]{Takata2006a}
{Takata}, M. 2006, \pasj, 58, 759

\bibitem[{{Takata}(2016{\natexlab{a}})}]{Takata2016a}
{Takata}, M. 2016{\natexlab{a}}, \pasj, 68, 109

\bibitem[{{Takata}(2016{\natexlab{b}})}]{Takata2016b}
{Takata}, M. 2016{\natexlab{b}}, \pasj, 68, 91

\bibitem[{{Tassoul}(1980)}]{Tassoul1980}
{Tassoul}, M. 1980, \apj, 43, 469

\bibitem[{{Unno} {et~al.}(1989){Unno}, {Osaki}, {Ando}, {Saio}, \&
  {Shibahashi}}]{Unno1989}
{Unno}, W., {Osaki}, Y., {Ando}, H., {Saio}, H., \& {Shibahashi}, H. 1989,
  {Nonradial oscillations of stars}

\bibitem[{{Vrard} {et~al.}(2016){Vrard}, {Mosser}, \& {Samadi}}]{Vrard2016}
{Vrard}, M., {Mosser}, B., \& {Samadi}, R. 2016, \aap, 588, A87

\end{thebibliography}

 \appendix

 \section{Value of $\theta_{\rm G}$ in the non-Cowling case}
 \label{phase lag}
 
Following \citet[][hereafter, T16]{Takata2016a}, the dependent variable $Y_1$ inside the buoyancy cavity is provided by Eqs.~(16), (114), and (116) of his paper. The dependent variable $Y_1$ reads in the asymptotic limit
 \algn{
 Y_1 (r) \sim A(r) \cos \left( \frac{\sqrt{2}}{\sigma} \int_0^{r} \frac{N}{r} \dd r\right) \; ,
 \label{A1}
 }
 where $A(r) $ is a real slowly varying amplitude. According to Eq.~(112) of T16, we can rewrite the phase in the cosine function of \eq{A1} such as
 \algn{
 \frac{\sqrt{2}}{\sigma} \int_0^{r} \frac{N}{r} \dd r \approx \int_{r_1}^{r} \mathcal{K}_r \dd r + \frac{5 \pi}{4} \; ,
 \label{A2}
 }
where $r_1$ is the innermost turning point. Injecting \eq{A2} into \eq{A1}, we obtain
 \algn{
 Y_1 \sim\frac{A(r)}{2} \left(\underbrace{e^{5i\pi/4} e^{i\Psi(r)}}_{\rm Downward~incident~energy~ray}+ \underbrace{e^{-5i\pi/4} e^{-i\Psi(r)}}_{\rm Upward~reflected~energy~ray}\right) \; ,
 \label{A3}
 }
 where the quantity
 \algn{
 \Psi(r) = \int_{r_1}^{r} \mathcal{K}_r \dd r \; 
 }
represents the rapidly varying wave phase that increases with respect to radius. A time dependence of $e^{-i\sigma t}$ is assumed for the oscillation variables in T16 and in this study. Therefore, the first (second) term inside the brackets of \eq{A3} corresponds to the progressive (regressive) component. Since the group velocity is in the opposite direction of the phase velocity in the buoyancy cavity, it is related to a downward incident (upward reflected) energy ray. Within the framework of the formulation of mixed modes by \cite{Takata2016b}, the phase lag $\theta_{\rm G}$ at the reflection in $r_1$ for the variable $Y_1$ is thus given by the argument of the ratio of the reflected to the incident wave amplitude in the sense of the group velocity, that is,
\algn{
\theta_{\rm G} = \arg \left( \frac{A(r) e^{-5i\pi/4}}{ A(r) e^{5i\pi/4}} \right) = - \frac{5 \pi}{2} \; .
}
In the Cowling approximation, the variable $Y_1$, which is equivalent to the variable $\varv$ in \cite{Shibahashi1979}, takes in contrast the form of an Airy function of the first kind near $r_1$. Following the same reasoning as the previous paragraphs, we can show that $\theta_{\rm G}=+\pi/2$ in this case (e.g., see discussion in \sectionname{}~\ref{general origin}).

 \section{Evolution of $\mathcal{N}_{\rm b}$ with $\nu_{\rm max}$ before the luminosity bump}
 \label{Nb evol}

The parameter $\mathcal{N}_{\rm b}$ represents the value of the modified Brunt-Väisälä frequency just below the base of the convective zone. Its value is well defined only in the ideal case where $\mathcal{N}$ varies as a power law in the radiative zone and discontinuously drops to about zero at the base of the convective zone, as schematically represented in \figurename{}~\ref{scheme}. However, as mentioned in \sectionname{}~\ref{discussion assumption}, stellar models suggest that $\mathcal{N}$ vanishes more progressively at the base of the convective zone so that a clear definition of this parameter can be subject to discussions \citep[e.g., see an attempt of definition in][]{Pincon2017}. Moreover, the transition from an adiabatic to a radiative temperature gradient at the base of the convective zone, which is related to the profile of $\mathcal{N}$ in this region, is still subject to uncertainties in stellar modeling. For these reasons, we choose to discuss the variation of this parameter during evolution using orders of magnitude through the proxy $\widetilde{N}_{\rm b}$, defined by
\algn{
\widetilde{N}_{\rm b} \equiv \frac{G m(r_{\rm b})}{r_{\rm b}^3} \; .
\label{Nb tilde}
}

\begin{figure}
\centering
\includegraphics[scale=0.35,trim= 0cm 0cm 0cm 0cm, clip]{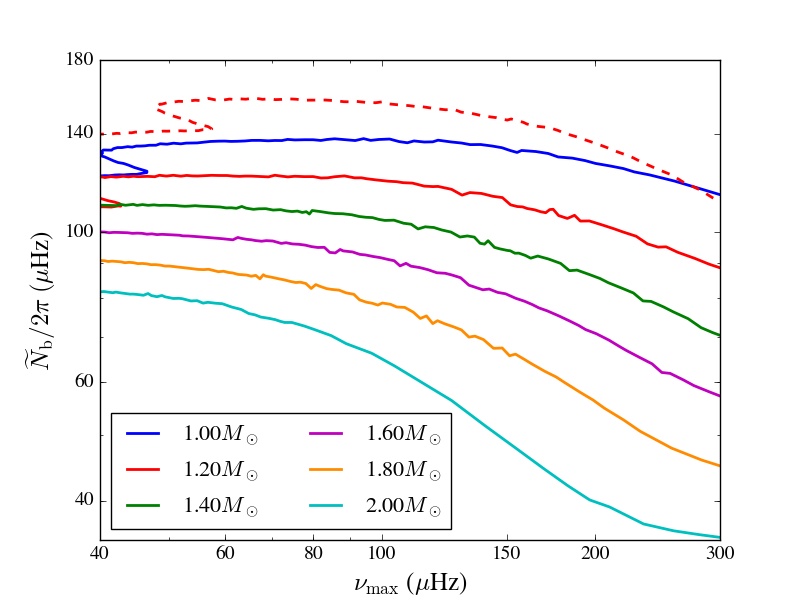} 
\caption{Evolution of $\widetilde{N}_{\rm b}$ defined in \eq{Nb tilde} as a function of $\nu_{\rm max}$ for standard evolutionary sequences of models with different masses (solid lines). The red dashed line represents the result for a similar $1.2M_\odot$ evolutionary sequence, except that a $0.3H$-thick adiabatic overshooting region was included at the base of the convective zone.} 
\label{N_b fig}
\end{figure}

The evolution of $\widetilde{N}_{\rm b}$ as a function of $\nu_{\rm max}$ is plotted in \figurename{}~\ref{N_b fig} for several evolutionary sequences of models with masses between $1M_\odot$ and $2M_\odot$. Most of these evolutionary sequences were built using the same standard input physics as used to build the $1.2M_\odot$ models considered as illustration in \sectionname{}~\ref{unclear}. Only one of these models, which has a mass of $1.2M_\odot$, includes adiabatic overshooting below the convective zone. The thickness of the overshooting region was assumed to be equal to about 0.3$H$, with $H$ the pressure scale height at the base of the convective zone, which is close to the mean value derived by \cite{Khan2018} to match the observed location of the luminosity bump in low-mass stars. The considered frequency range for $\nu_{\rm max}$ correspond to stars between the beginning of the RGB and the luminosity bump. In \figurename{}~\ref{N_b fig}, we see that \smash{$\widetilde{N}_{\rm b}$} generally increases as $\nu_{\rm max}$ decreases at this evolutionary stage, except for a short phase before the luminosity bump (corresponding to the rapid variation visible at low values of $\nu_{\rm max}$ in \figurename{}~\ref{N_b fig}). Indeed, on the first part of the RGB, the base of the convective zone progressively deepens into the interior of the stars with respect to both the mass and radius coordinates. At a certain stage just before the luminosity bump, it attains a minimum and starts slightly receding. This result in the first dredge-up of the chemical elements \citep[e.g.,][]{Kippenhahn2012}. Since \smash{$\widetilde{N}_{\rm b}\propto m(r_{\rm b})/r_{\rm b}^3$}, this parameter is more sensitive to the decrease in $r_{\rm b}$ than in $m(r_{\rm b})$ and thus increases in the first part of the RGB, as depicted in \figurename{}~\ref{N_b fig}. In the short phase just after the convective boundary has attained its minimum and before the luminosity bump, \smash{$\widetilde{N}_{\rm b}$} decreases so slowly that it can be regarded as constant at first approximation. We also see that the inclusion of an adiabatic overshooting region at the base of the convective region significantly modifies the value of \smash{$\widetilde{N}_{\rm b}$}. Indeed, in this case, the radius of the base of the convective region is smaller than without overshooting while its mass coordinate remains about constant, so that \smash{$\widetilde{N}_{\rm b}$} must be higher for the same reasons as before. Nevertheless, we see in \figurename{}~\ref{N_b fig} that the variation rate of \smash{$\widetilde{N}_{\rm b}$} as a function of $\nu_{\rm max}$ remains comparable during evolution with or without overshooting. Moreover, we checked by varying the overshooting length that the thicker the overshooting region, the higher \smash{$\widetilde{N}_{\rm b}$}, as expected from the previous arguments. 

\begin{figure}
\centering
\includegraphics[scale=0.35,trim= 0cm 0cm 0cm 0cm, clip]{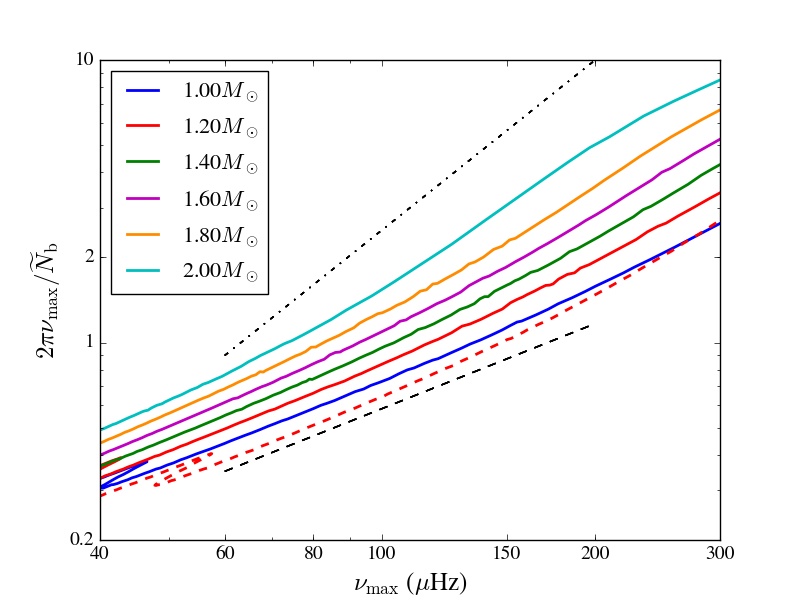} 
\caption{Same caption as in \figurename{}~\ref{N_b fig} but for the ratio $2\pi \nu_{\rm max} /\widetilde{N}_{\rm b}$. The black dashed and dash-dotted lines symbolize the slope in the cases where $x=0$ or $x=1$ in \eq{ratio assumption}. The red dashed line has the same definition as in \figurename{}~\ref{N_b fig}} 
\label{ratio}
\end{figure}

In \figurename{}~\ref{ratio}, we see that \smash{$\widetilde{N}_{\rm b}$} increases more slowly than $\nu_{\rm max}$ decreases during evolution. This is because the stellar radius increases more rapidly than $r_{\rm b}$ decreases at these stages. As a result, it is obvious that the ratio \smash{$2\pi \nu_{\rm max} /\widetilde{N}_{\rm b}$} decreases during evolution more slowly than $\nu_{\rm max}^2$, as assumed in \sectionname{}~\ref{comparison}. Figure~\ref{ratio} also shows that at a given value of $\nu_{\rm max}$, the higher the mass, the higher this ratio. This confirms the fact the higher the mass, the lower the value of $\nu_{\rm max,t}$ at the transition between cases $a$ and $b$ (i.e., when \smash{$2\pi \nu_{\rm max} \approx \mathcal{N}_{\rm b} \sim \widetilde{N}_{\rm b}$)}. Moreover, since the thicker the overshooting region, the higher \smash{$\widetilde{N}_{\rm b}$} at a given value of $\nu_{\rm max}$, we deduce that the lower the ratio \smash{$2\pi \nu_{\rm max} /\widetilde{N}_{\rm b}$} so that the higher the value of $\nu_{\rm max,t}$ at the transition, as shown in \figurename{}~\ref{ratio}.

In summary, considering that \smash{$\widetilde{N}_{\rm b}\sim \mathcal{N}_{\rm b}$}, we conclude the following for stellar masses between $1M_\odot$ and $2M_\odot$:
\begin{itemize}
\item The value of $\mathcal{N}_{\rm b}$ increases at the beginning of the ascent of the RGB until a short phase just before the luminosity bump where it remains about constant.
\item The ratio \smash{$2\pi \nu_{\rm max} /\mathcal{N}_{\rm b}$} decreases more slowly than $\nu_{\rm max}^2$ during evolution since the envelope expansion is faster than the migration of the convective boundary toward the center.
\item The thicker the overshooting region below the convective region or the lower the stellar mass, the higher the values of $\mathcal{N}_{\rm b}$ and of $\nu_{\rm max,t}$.
\end{itemize}

 \section{Seismic properties of the four outliers}
 \label{outlier}
 
The seismic properties of the four outliers depicted in \figurename{}~\ref{obs} are summarized in \tablename{}~\ref{table2} following the data provided by \cite{Mosser2018}.
 
   \begin{table}[h]
     $$ 
         \begin{array}{cccccccc}
           \hline
          \hline
        \noalign{\smallskip}
            {\rm KIC ~ \#}& \nu_{\rm max}& \Delta \nu&\Delta \Pi_1&M &\nu_{\rm rot}&q&\epsilon_{\rm g}^{\rm obs}  \\
                \noalign{\smallskip}
        &[\mu{\rm Hz}]& [\mu{\rm Hz}]&[s]&[M_\odot]&[\mu{\rm Hz}]&\slash&\slash \\
        \noalign{\smallskip}
            \hline
        \noalign{\smallskip}
        2584478&182& 14&81&1.38&0.51&0.13&-0.14\\
        \noalign{\smallskip}
        12008916&161&13&79&1.33&0.52&0.15&0.017\\
        \noalign{\smallskip}
        3749487&70&6.6&68&1.55&0.49&0.12&-0.31\\
        \noalign{\smallskip}
        10387370&69&7.1&70&1.09&0.50&0.12&-0.42\\
        \noalign{\smallskip}
        \hline
         \end{array}
     $$ 
           \caption[]{Seismic properties of the four outliers discussed in \sectionname{}~\ref{comparison}. $\nu_{\rm rot}$ represents the core rotation frequency and $q$ is the coupling factor. The seismic masses are computed from the seismic scaling laws provided by \cite{Mosser2013b}.}
         \label{table2}
   \end{table}
%

\end{document}